\documentclass[twocolumn,preprintnumbers,amsmath,amssymb]{revtex4}
\usepackage{amsmath}
\usepackage{extarrows}
\usepackage[ruled]{algorithm2e}
\usepackage{graphicx}
\usepackage{dcolumn}
\usepackage{bm}
\usepackage[all]{xy}
\usepackage{indentfirst}
\usepackage{amsmath}
\usepackage{multirow}
\usepackage{mathrsfs}
\usepackage{bbm}
\usepackage{euscript}
\usepackage{amssymb}
\usepackage{extarrows}
\usepackage{bbm}
\usepackage[ruled]{algorithm2e}
\usepackage{graphicx}
\usepackage{epstopdf}
\usepackage{dcolumn}
\usepackage{bm}
\usepackage[all]{xy}
\usepackage{indentfirst}
\usepackage{amsmath}
\usepackage{multirow}
\usepackage{mathrsfs}
\usepackage{euscript}
\usepackage{amssymb}
\usepackage{appendix}
\usepackage{color,xcolor}

\begin{document}

\title{A new parameterized entanglement monotone}

\author{Xue Yang$^1$, Ming-Xing Luo$^1$, Yan-Han Yang$^1$, Shao-Ming Fei$^{2,3}$}

\affiliation{1. The School of Information Science and Technology, Southwest Jiaotong University, Chengdu 610031, China;
\\
2. School of Mathematical Sciences, Capital Normal University, Beijing 100048, China;
\\
3. Max-Planck-Institute for Mathematics in the Sciences, 04103 Leipzig, Germany}

\begin{abstract}
Entanglement concurrence has been widely used for featuring entanglement in quantum experiments. As an entanglement monotone it is related to specific quantum Tsallis entropy. Our goal in this paper is to propose a new parameterized bipartite entanglement monotone which is named as $q$-concurrence inspired by general Tsallis entropy. We derive an analytical lower bound for the $q$-concurrence of any bipartite quantum entanglement state by employing positive partial transposition criterion and realignment criterion, which shows an interesting relationship to the strong separability criteria. The new entanglement monotone is used to characterize bipartite isotropic states. Finally, we provide a computational method to estimate the $q$-concurrence for any entanglement by superposing two bipartite pure states. It shows that the superposition operations can at most increase one ebit for the $q$-concurrence in the case that the two states being superposed are bi-orthogonal or one-sided orthogonal. These results reveal a series of new phenomena about the entanglement, which may be interesting in quantum communication and quantum information processing.
\end{abstract}

\maketitle

\section{Introduction}

Quantum entanglement as one of the most remarkable phenomena of quantum mechanics, reveals the fundamental insights into the nature of quantum correlations. It is key of many interesting quantum tasks such as quantum teleportation \cite{Bennett1}, quantum dense coding \cite{Bennett2}, quantum secret sharing \cite{Hillery}, and quantum cryptography \cite{Gisin}. An fundamental problem is to justify whether a given quantum composite system state is entangled or separable. So far, there are two important entanglement criteria for the bipartite entanglement. One is positive partial transpose (PPT) criterion  \cite{Peres(1996)} which implies the partial transposition satisfying $\rho^{T_A}\geq 0$ for any separable state $\rho_{AB}$. The PPT criterion is a necessary and sufficient condition of entanglement for pure states, $2\otimes2$ and $2\otimes3$  mixed states, but in
general not sufficient for higher dimensions \cite{Peres(1996),Horodecki(1996)}. The other is complementary operational criterion which is called the realignment criterion \cite{Horodecki(1999),Rudolph (2005),Chen(2002)}. For a separable $\rho_{AB}$, the realignment operation ${\cal{R}}(\rho)$ satisfies $\|{\cal{R}}(\rho)\|_1\leq1$. Both entanglement criteria are widely used in quantum experiments and quantum applications \cite{3H(2009)}.

Entanglement measure as another approach is also used to quantify entanglement \cite{Vedral(1997),Vedral(1997)1}. There are some interesting entanglement measures for bipartite entangled systems, such as the concurrence \cite{Hill(1997),Wootters(1998),Rungta(2001)}, entanglement of formation \cite{Bennett(1996),Horodeck(2001)}, negativity  \cite{Zyczkowski(1998),Vidal(2002)}, Tsallis-$q$ entropy of entanglement  \cite{Kim(2010)T}, and R\'{e}nyi-$\alpha$ entropy of entanglement \cite{Gour(2007),Kim(2010)R}. Howbeit, the explicit computation of these measures for arbitrary states is a formidable task because of the extremization for mixed states. So far,  analytical results are only available for special measures and two-qubit states or special higher-dimensional mixed states \cite{Lee(2003),Wootters(1998),Rungta(2003),Vollbrecht(2001),Terhal(2000),Buchholz(2016)}. Moreover, these entanglement measures are also related to the PPT criterion and the realignment criterion. Recently, some efforts have been made towards the analytical lower bounds of concurrence \cite{Chen(2005), L.G.Liu(2009),M. Li(2018)}.  Thereby, the development of analytical lower bounds for the various entanglement measures is of great interest.

In fact, as one reason of the difficult quantification of mixed states, any entanglement generated by superposing two pure states cannot be simply featured by two individual states being superposed. Entanglement of superpositions is firstly introduced by Linden et al. \cite{Linden(2006)} who found an upper bound on the entanglement of formation for the superposition in terms of the entanglement of two individual states being superposed. Their bound is then improved by  Gour \cite{Gour(2007)1}. So far, the entanglement of superpositions has been addressed  in terms of different entanglement measures \cite{K.H. Ma(2010),Akhtarshenas(2011),Yu(2006),Song(2007),Xiang(2007),Ou(2007),Ma(2014)}. Although it is difficult to exactly estimate the entanglement measure of the superposed entanglement from individual states being superposed, however, it may be helpful for exploring the nature of quantum entanglement by investigating the superposed entanglement. Additionally, it provides direct results for the approximate quantification of mixed entanglement \cite{Lohmayer(2006),Osterloh(2008),Eltschka(2008)}.

The concurrence for a bipartite pure state $|\psi\rangle_{AB}$ is defined by $C(|\psi\rangle_{AB})=\sqrt{2(1-{\rm{Tr}}\rho^2_A)}$  \cite{Hill(1997)}. It plays a major role in entanglement distributions such as entanglement swapping and remote preparation of bipartite entangled states \cite{Gour(2004)}. In fact, the concurrence for pure states is related to specific Tsallis entropy  \cite{Tsallis(1988),Landsberg(1998)} as $C(|\psi\rangle_{AB})=\sqrt{2T_2(\rho_A})$ for $q=2$. Noteworthily, Tsallis entropy provides a generalization of traditional Boltzmann-Gibbs statistical mechanics and  enables us to find a consistent treatment of dynamics in many nonextensive physical systems such as long-range interactions, long-time memories, and multifractal structures \cite{Tsallis(2001)1}. Tsallis entropy also provides many intriguing applications in the realms of quantum information theory \cite{Abe(2001),Tsallis(2001),Rossignoli(2002),Rajagopal (2005)}. Hence, a natural problem is how to construct an entanglement measure from general Tsallis entropy with $q\geq2$. Our goal in this paper is to solve this problem.

The outline of the rest is as follows. In Sec.II, we propose a new bipartite entanglement monotone which is related to general Tsallis entropy for any $q\geq 2$. The so-called $q$-concurrence is actually an entanglement monotone. We prove an analytical lower bound for the $q$-concurrence by using the PPT and realignment criteria. Moreover, we valuate  the $q$-concurrence for isotropic states. In Sec.III, we investigate the entanglement of the superposition of two pure states by using the $q$-concurrence in terms of two states being superposed. The entanglement of the superposed state can be expressed explicitly when two input states are bi-orthogonal or one-sided orthogonal. As a result, the superposing operation can only increase at least one ebit in terms of the $q$-concurrence in both cases. The last section concludes the paper.

\section{A new entanglement monotone}

Before giving our definition, we recall a well-known bipartite entanglement monotone. For any arbitrary bipartite pure state $|\psi\rangle_{AB}$ on Hilbert space ${\cal H}_A\otimes {\cal H}_{B}$, the concurrence \cite{Hill(1997),Rungta(2001)} is given by
\begin{eqnarray}
C(|\psi\rangle_{AB})=\sqrt{2(1-{\rm{Tr}}\rho^2_A)}
\label{eqn01}
\end{eqnarray}
where $\rho_A={\rm{Tr}}_B(|\psi\rangle_{AB}\langle\psi|)$ is the reduced density matrix of the subsystem $A$ by tracing out the subsystem $B$.

The concurrence defined in Eq.(\ref{eqn01}) can be regarded as a function of specific Tsallis entropy of $q=2$ \cite{Tsallis(1988),Landsberg(1998)}, i.e., $C(|\psi\rangle_{AB})=\sqrt{2 T_2(\rho_A})$ for $q=2$. In this section, we define another new parameterized entanglement monotone named the $q$-concurrence which is related to general Tsallis entropy for any $q\geq2$.

\subsection{The $q$-concurrence}

\textbf{Definition 1.} For an arbitrary bipartite pure state $|\psi\rangle_{AB}$ on Hilbert space ${\cal H}_A\otimes {\cal H}_{B}$, the $q$-concurrence is defined as
\begin{eqnarray}
C_q(|\psi\rangle_{AB})=1-{\rm{Tr}}\rho^q_{A}
\label{eqn0}
\end{eqnarray}
for any $q\geq2$, where $\rho_{A}$ is the reduced density operator of the subsystem $A$.

It is clear that $C_q(|\psi\rangle_{AB})=0$ if and only if $|\psi\rangle_{AB}$ is a separable state, i.e., $|\psi\rangle_{AB}=|\psi\rangle_{A}\otimes|\psi\rangle_{B}$. The $q$-concurrence may be concerned with Schatten $q$-norm for the positive semidefinite matrices, where the Schatten $q$-norm \cite{Bhatia(1997)} is defined as
\begin{eqnarray}
\|A\|_q=({\rm Tr}A^q)^{1/q}
\end{eqnarray}
It will be a useful tool to prove the subadditivity inequality in the following Lemma 1.

Suppose a pure state $|\psi\rangle_{AB}$ defined on Hilbert space ${\cal H}_A\otimes {\cal H}_{B}$ has the Schmidt decomposition
\begin{eqnarray}
|\psi\rangle=\sum^m_{i=1}\sqrt{\lambda_i}|a_i\rangle_A|b_i\rangle_B
\label{eqnSchmidt1}
\end{eqnarray}
It is apparent that the reduced density matrices $\rho_A$ and $\rho_B$ have the same spectra of $\{\lambda_i\}$. Hence, we have
\begin{eqnarray}
C_q(|\psi\rangle_{AB})=1-{\rm{Tr}}\rho^q_{A}=1-{\rm{Tr}}\rho^q_{B}
\end{eqnarray}
This implies that
\begin{eqnarray}
C_q(|\psi\rangle)=1-\sum^m_{i=1} {\lambda^q_i}
\label{eqni}
\end{eqnarray}
where $C_q(|\psi\rangle)$ satisfies $0\leq C_q(|\psi\rangle)\leq 1-m^{1-q}$.  The lower bound is obtained for product states while the upper bound is achieved for the maximally entangled pure states $\frac{1}{\sqrt{m}}\sum^m_{i=1} |ii\rangle$.

For a mixed state $\rho_{AB}$ on Hilbert space ${\cal H}_A\otimes {\cal H}_{B}$, we define its $q$-concurrence via the convex-roof extension as follows:
\begin{eqnarray}
C_q(\rho_{AB})=\inf_{\{p_i,|\psi_i\rangle\}}\sum_ip_iC_q(|\psi_i\rangle_{AB})
\label{eqnmixed}
\end{eqnarray}
where the infimum is taken over all the pure-state decompositions of $\rho_{AB}=\sum_ip_i|\psi_i\rangle\langle\psi_i|$ with $\sum_ip_i=1$, and $p_i\geq0$.

So far, several results have been made for the requirements that a reasonable measure of entanglement should fulfill \cite{Popescu(1997),Vedral(1997)1,Vidal(1999)}. Specially, it has been proposed in \cite{Vidal(2000)} that the monotonicity under local operations and classical communication (LOCC) has to satisfy as the only requirement of any entanglement measure. This kind of entanglement measure is then defined as entanglement monotone. In fact, Vidal \cite{Vidal(2000)} states that it is an entanglement monotone $E$  if the following conditions hold:
\begin{itemize}
\item[(i)] $E(\rho)\geq 0$ for any state $\rho$, and $E(\rho)=0$ if $\rho$ is fully separable;
\item[(ii)] For a pure state $|\Psi\rangle$, the measure is a function of
the reduced density operator $\rho_A={\rm{Tr}}_B(|\Psi\rangle\langle\Psi|)$, i.e., $E(|\Psi\rangle)=f(\rho_A)$, where the function $f$ has the following properties: (a) $f$ is invariant under any unitary transformation $U$, i.e. $f(U\rho_AU^\dagger)=f(\rho_A)$. (b) $f$ is concave, i.e., $f(\lambda\rho_1+(1-\lambda)\rho_2)\geq \lambda f(\rho_1)+(1-\lambda) f(\rho_2)$ for  $\lambda\in (0,1)$;
\item[(iii)] For a mixed state $\rho$, the measure $E(\rho)$ is defined as the convex-roof extension, i.e.,
\begin{eqnarray}
E(\rho)=\inf_{\{p_i,|\psi_i\rangle\}}
\{\sum_ip_iC(|\psi_i\rangle)|\sum_ip_i|\psi_i\rangle\langle\psi_i|=\rho\}
\end{eqnarray}
where the minimum is taken over all possible pure-state decompositions of $\rho$.
\end{itemize}

These conditions (i)-(iii) formalize intuitive properties of an entanglement monotone.  From this point of view any entanglement monotone could be regarded as a measure of entanglement. We present the following Lemma 1 for verifying that the $q$-concurrence defined in Eq.(\ref{eqnmixed}) is a proper entanglement monotone.

\textbf{Lemma 1}. Define the function
\begin{eqnarray}
F_q(\rho)=1-{\rm{Tr}}\rho^q
\label{eqnFq}
\end{eqnarray}
for any density matrix $\rho$ and $q\geq2$. $F_q(\rho)$ satisfies the following properties.
\begin{itemize}
\item[(i)] \textbf{Nonnegativity}: $F_q(\rho)\geq 0$ for any density operator $\rho$,  where the equality holds for pure states;
\item[(ii)] \textbf{Symmetry}: $F_q(\rho_A)=F_q(\rho_B)$ for a  pure state $\rho_{AB}$ of the composite system $AB$;
\item[(iii)]  \textbf{Subadditivity}: For a general bipartite state $\rho_{AB}$,   $F_q(\rho_{AB})$ satisfies the inequalities:
\begin{eqnarray}
|F_q(\rho_A)-F_q(\rho_B)|\leq F_q(\rho_{AB})\leq F_q(\rho_A)+F_q(\rho_B)
\end{eqnarray}
\item[(iv)] \textbf{Concavity and quasi-convexity}: $F_q$ is concave, i.e.,
\begin{eqnarray}
\sum_i p_iF_q(\rho_i)\leq F_q(\sum_i p_i\rho_i)
\label{eqntriangle}
\end{eqnarray}
where $\{p_i\}$ is a probability distribution, and $\rho_i$s are density matrices. The equality holds iff $\rho_i$s are identical for all $p_i>0$. Moreover,  $F_q$ is  quasi-convex, i.e.,
\begin{eqnarray}
F_q(\sum_i p_i\rho_i)\leq\sum_i p^q_iF_q(\rho_i)+1-\sum_i p^q_i
\label{eqntriangle}
\end{eqnarray}
where the equality holds iff $\rho_i$ have supports on orthogonal subspaces, i.e., $\rho_i=|\psi_i\rangle\langle\psi_i|$, and $\{|\psi_i\rangle\}$ are orthogonal.
\end{itemize}

The proof of the Lemma 1 is provided in Appendix A. Next we prove $C_q(\rho)$ is a proper entanglement monotone.

\textbf{Proposition 1}. The $q$-concurrence $C_q(\rho)$ in Eq.(\ref{eqnmixed}) is an entanglement monotone.

\textbf{Proof.} From the non-negativity in Lemma 1 and Eq.(\ref{eqnmixed}), it follows that $C_q(\rho)\geq 0$ for any density matrix $\rho$, where the equality holds iff $\rho$ is separable. Furthermore, from the concavity in Lemma 1, we know that $F_q(\lambda\rho_1+(1-\lambda)\rho_2)\geq \lambda F_q(\rho_1)+(1-\lambda) F_q(\rho_2)$ for any density matrices $\rho_1$ and $\rho_2$ and $\lambda\in (0,1)$. Thus, $C_q(|\phi\rangle_{AB})$ is a concave function of $\rho_A$. Finally,  $C_q(|\phi\rangle_{AB})$ is invariant under local unitary transformations from the invariance of ${\rm{Tr}}\rho^q$.  Then, the convex-roof extension of the $q$-concurrence $C_q(\rho)$ for mixed states is a proper entanglement monotone \cite{Vidal(2000)}. $\Box$

Note that for a given bipartite entanglement $|\psi\rangle$, the entanglement monotone  $C_q(|\psi\rangle)$ in Eq.(\ref{eqn0}) are invariant under the local unitary operations. From Cayley-Hamilton Theorem, the reduced density matrix $\rho_A$ of rank $d$ satisfies a characteristic equation as $\sum_{j=0}^da_j\rho_A^j=0$. From the spectra decomposition of $\rho_A=\sum^d_{i=1}\lambda_i|\phi_i\rangle\langle \phi_i|$, from Eq.(\ref{eqni}) the $q$-concurrence satisfies a linear equation as $\sum_{q=0}^da_qC_q(|\psi\rangle)-\sum_ia_iC_0(|\psi\rangle)=0$. Hence, all the $q$-concurrences with $q\leq d$ will be evaluated for any $d$-dimensional pure states.  However, for a mixed state $\rho_{AB}$ the $q$-concurrences $C_q(\rho_{AB})$ in Eq.(\ref{eqnmixed}) do not satisfy the characteristic equation of the density matrix $\rho_A$, or $\rho_{AB}$. This implies that each $q$-concurrence of $C_q(\rho_{AB})$ may provide different meanings for featuring the entanglement. In fact, as a parameterized generalization of the Boltzmann-Gibbs entropy, the Tsallis entropy is specially interesting in long-range systems such as the motion of cold atoms in dissipative optical lattices \cite{Lutz,Douglas}, spin glass relaxation \cite{Pickup} or trapped ion \cite{Devoe}. The present $q$-concurrences for general $q\geq2$ may be applicable for featuring these long-range entangled systems beyond the standard concurrence \cite{Hill(1997)}.

\subsection{A lower bound on the $q$-concurrence}

In contrast to the simple case of pure entangled states in Eq.(\ref{eqn0}), the quantification of mixed states is still challenging due to the optimization procedures \cite{3H(2009)}. Fortunately, we present an effective operational way to detect the $q$-concurrence for any bipartite quantum state, which manifests an essential quantitative relation among the $q$-concurrence, PPT criterion and realignment criterion.  Before presenting the lower bound, we recall two separability criteria.

\emph{PPT criterion}\cite{Peres(1996),Horodecki(1996)}. Given a bipartite state $\rho_{AB}=\sum_{ijkl}\rho_{ij,kl}|ij\rangle\langle kl|$. If $\rho_{AB}$ is separable, then the partial transposition $\rho^{T_A}$ with respect to the subsystem $A$ has the nonnegative spectrum, i.e., $\rho^{T_A}\geq 0$. The partial transpose $\rho^{T_A}$ is given by $\rho^{T_A}=[\sum_{ijkl}\rho_{ij,kl}|ij\rangle\langle kl|]^{T_A}=\sum_{ijkl}\rho_{ij,kl}|kj\rangle\langle il|$, where the subscripts $i$ and $j$ are the row and column indices for the subsystem $A$, respectively, while $k$ and $l$ are such indices for the subsystem $B$.

\textit{Realignment criterion} \cite{Horodecki(1999),Rudolph (2005),Chen(2002)}. Let ${\cal{R}}$ be the realignment operation on the joint system $\rho_{AB}=\sum_{ijkl}\rho_{ij,kl}|ij\rangle\langle kl|$. The output is given by ${\cal{R}}(\rho)=\sum_{ijkl}\rho_{ij,kl}|ik\rangle\langle jl|$.  If $\rho_{AB}$ is separable, then $\|{\cal{R}}(\rho)\|_1\leq1$, where $\|X\|_1$ denotes the trace norm defined by $\|X\|_1=\rm{Tr}\sqrt{XX^\dagger}$ \cite{Rudolph(2004)}.

 According to these criteria, a given state $\rho$ is entangled if the trace norms $\|\rho^{T_A}\|_1$  or $\|{\cal{R}}(\rho)\|_1$ are strictly larger than 1.

 For the pure state defined in Eq.(\ref{eqnSchmidt1}), it is straightforward to prove \cite{K.Chen(2005)}:
\begin{eqnarray}
\| \rho^{T_A}\|_1=\|{\cal{R}}(\rho)\|_1=(\sum^m_{i=1} \sqrt{\lambda_i})^{2}
\label{eqnTR}
\end{eqnarray}
Besides, we know that
\begin{eqnarray}
(\sum^m_{i=1} \sqrt{\lambda_i})^{2}\leq  m\sum^m_{i=1} \lambda_i=m
\label{eqnTR1}
\end{eqnarray}
from the Cauchy-Schwarz inequality, where $\{\lambda_i\}$ is a probability distribution.

\textbf{Theorem 1}. For any mixed entanglement state $\rho$ on Hilbert space ${\cal H}_A\otimes {\cal H}_{B}$ with the dimension of $m$ and $n$ ($m \leq n$), respectively, the $q$-concurrence $C_q(\rho)$ satisfies the following inequality
\begin{eqnarray}
C_q(\rho)\geq\frac{(\max{\{\|\rho^{T_A}\|_1^{q-1},\|{\cal{R}}(\rho)\|_1^{q-1}\}}-1)^2}{m^{2q-2}-m^{q-1}}
\label{eqnmixed0}
\end{eqnarray}

\textbf{Proof}. Consider the optimal decomposition of $\rho$ as $\rho=\sum_i p_i \rho_i $ in order to achieve the infimum of $C_q(\rho)$ in Eq.(\ref{eqnmixed}), where $\rho_i$s are pure states with $\rho_i=|\psi_i\rangle\langle\psi_i|$. Firstly, we will prove that
\begin{eqnarray}
C_q(\rho_i)&\geq&\frac{(\|\rho_i^{T_A}\|_1^{q-1}-1)^2}{m^{2q-2}-m^{q-1}}
\label{eqnTa}
\\
C_q(\rho_i)&\geq&\frac{(\| {\cal{R}}(\rho_i)\|_1^{q-1}-1)^2}{m^{2q-2}-m^{q-1}}
\label{eqnRa}
\end{eqnarray}

In fact, note that the function $g(\lambda)=\lambda^q$ is convex for $q\geq 2$ and $\lambda\in (0,1)$. It means that $h(\lambda_1, \cdots, \lambda_m)=\sum_{k=1}^m\lambda_k^q$ is Schur-convex \cite{Bhatia(1997)}. Since the uniform distribution of $\{\frac{1}{m}, \cdots, \frac{1}{m}\}$ is majorized by any other distribution $\{\lambda_1, \cdots, \lambda_m\}$, i.e., $\{\frac{1}{m}, \cdots, \frac{1}{m}\}\prec \{\lambda_1, \cdots, \lambda_m\}$. For the Schur-convex function $h(\lambda_1, \cdots, \lambda_m)$ it follows that \cite{Bhatia(1997)}:
\begin{eqnarray}
\sum^m_{k=1} \lambda^q_k\geq \frac{1}{m^{q-1}}
\label{Schur}
\end{eqnarray}
From Eq.(\ref{eqni}), we have
\begin{eqnarray}
C_q(\rho_i)
 &=&1-\sum^m_{k=1} \lambda_{ik}^q
 \nonumber \\
 &=&\frac{(\sum^m_{j=1} \sqrt{\lambda_{ij}})^{2q-2}-\sum^m_{k=1} \lambda_{ik}^q(\sum^m_{j=1} \sqrt{\lambda_{ij}})^{2q-2}}{(\sum^m_{j=1} \sqrt{\lambda_{ij}})^{2q-2}}
\nonumber \\
 &\geq&\frac{(\sum^m_{k=1}\sqrt{\lambda_{ik}})^{2q-2}-1}{(\sum^m_{k=1} \sqrt{\lambda_{ik}})^{2q-2}}
\label{eqnpure0}
 \\
&=&\frac{((\sum^m_{j=1} \sqrt{\lambda_{ij}})^{2q-2}-1)^2}{(\sum^m_{k=1} \sqrt{\lambda_{ik}})^{2q-2}((\sum^m_{k=1} \sqrt{\lambda_{ik}})^{2q-2}-1)}
\nonumber \\
 &\geq &\frac{(\|\rho_i^{T_A}\|_1^{q-1}-1)^2}{m^{2q-2}-m^{q-1}}
\label{eqnpure2}
\end{eqnarray}
where the inequality (\ref{eqnpure0}) is due to the inequality: $-\sum^m_{k=1} \lambda_{ik}^q(\sum^m_{j=1} \sqrt{\lambda_{ij}})^{2q-2}\geq-1$ which can be proved by using the inequalities (\ref{eqnTR1}) and (\ref{Schur}). Moreover, for a pure state $\rho_i=|\psi_i\rangle\langle\psi_i|$, we have $\|\rho_i^{T_A}\|_1=(\sum^m_{k=1} \sqrt{\lambda_{ik}})^{2}\leq m$ as shown in Eq.(\ref{eqnTR}). This implies the inequality (\ref{eqnpure2}).

From Eq.(\ref{eqnpure2}), we have
\begin{eqnarray}
\sum_i p_iC(\rho_i)\geq \frac{\sum_i p_i(\|\rho_i^{T_A}\|_1^{q-1}-1)^2}{m^{2q-2}-m^{q-1}}
\label{eqnmixed1}
\end{eqnarray}
In what follows, we prove that
\begin{eqnarray}
(\|\rho^{T_A}\|_1^{q-1}-1)^2\leq\sum_i p_i(\|\rho_i^{T_A}\|_1^{q-1}-1)^2
\end{eqnarray}
In fact, for $\rho^{T_A}= \sum_i p_i\rho_i^{T_A}$, we obtain
\begin{eqnarray}
(\|\rho^{T_A}\|_1^{q-1}-1)^2
 \nonumber &=&(\| \sum_i p_i\rho_i^{T_A}\|_1^{q-1}-1)^2
 \\&\leq &(\sum_i p_i\| \rho_i^{T_A}\|_1^{q-1}-1)^2
\label{eqnfs1}
\\ \nonumber &\leq &\sum_i p_i\|\rho_i^{T_A}\|_1^{2q-2}
\\&&-2\sum_i p_i\|\rho_i^{T_A}\|_1^{q-1}+1
\label{eqnfs2}
 \\&=&\sum_i p_i(\|\rho_i^{T_A}\|_1^{q-1}-1)^2
\label{eqnfs}
\end{eqnarray}
Note that we have $\|\sum_ip_i\rho_i^{T_A}\|_1^{q-1}\leq \sum_ip_i\|\rho_i^{T_A}\|_1^{q-1}$ from the convexity of function $f(x)=\|x\|_1^{q-1}$ with $q\geq2$. Moreover, $\|\sum_ip_i\rho_i^{T_A}\|_1^{q-1}\geq 1$ and $\sum_ip_i\|\rho_i^{T_A}\|_1^{q-1}\geq1$ from $\|\rho_i^{T_A}\|_1^{q-1}\geq1$ for any density matrix $\rho_i$ and $q\geq 2$. This follows the inequality (\ref{eqnfs1}). The inequality (\ref{eqnfs2}) is obtained from the convexity of the function $f(x)=x^{2}$, i.e.,
\begin{eqnarray}
(\sum_i p_i\|\rho_i^{T_A}\|_1^{q-1})^2\leq\sum_i p_i\|\rho_i^{T_A}\|_1^{2q-2}
\label{}
\end{eqnarray}
Thereby, by substituting Eq.(\ref{eqnfs}) into Eq.(\ref{eqnmixed1}) we obtain
\begin{eqnarray}
C_q(\rho)\geq\frac{(\|\rho^{T_A}\|_1^{q-1}-1)^2}{m^{2q-2}-m^{q-1}}
\label{eqnTA}
\end{eqnarray}

From Eq.(\ref{eqnTR}), similar to Eq.(\ref{eqnpure2}) we can prove that
\begin{eqnarray}
C_q(\rho)\geq\frac{(\| {\cal{R}}(\rho)\|_1^{q-1}-1)^2}{m^{2q-2}-m^{q-1}}
\label{eqnRA}
\end{eqnarray}

Combining Eqs.(\ref{eqnTA}) and (\ref{eqnRA}), we get that
\begin{eqnarray}
C_q(\rho)\geq\max\{\frac{(\|\rho^{T_A}\|_1^{q-1}-1)^2}{m^{2q-2}-m^{q-1}},
\frac{(\| {\cal{R}}(\rho)\|_1^{q-1}-1)^2}{m^{2q-2}-m^{q-1}}\}
\label{eqnss}
\end{eqnarray}
Note that $\|\rho^{T_A}\|_1^{q-1}\geq 1 $ and $\| {\cal{R}}(\rho)\|_1^{q-1}\geq 1$ for any density matrix $\rho$ and $q\geq 2$. It follows the inequality (\ref{eqnmixed0}) from the inequality (\ref{eqnss}). This completes the proof. $\square$

\textit{Example 1}. Isotropic states are a class of mixed states on $\mathbb{C}^d\otimes \mathbb{C}^d$ which are invariant under the operation $U\otimes U^\ast$ with any unitary transformation $U$. Such mixed states are generally expressed as \cite{Horodecki(1999)}:
\begin{eqnarray}
\rho_F=\frac{1-F}{d^2-1}(\mathbbm{1}-|\Psi^{+}\rangle\langle\Psi^{+}|)+F|\Psi^{+}\rangle\langle\Psi^{+}|
\end{eqnarray}
where $\mathbbm{1}$ denotes the identity operator, $|\Psi^{+}\rangle=\frac{1}{\sqrt{d}}\sum^d_{i=1}|ii\rangle$ and $F$ is the fidelity of $\rho_F$ with respect to $\rho_{\Psi}=|\Psi^{+}\rangle\langle\Psi^{+}|$, i.e., $F=f_{\Psi^{+}}(\rho_F)=\langle\Psi^{+}|\rho_F|\Psi^{+}\rangle$ which satisfies $0\leq F\leq1$.

For isotropic states $\rho_F$, there are lots of analytic results in the entanglement of formation \cite{Terhal(2000)}, the tangle and concurrence \cite{Rungta(2003)}, and R\'{e}nyi $\alpha$-entropy entanglement \cite{Wang(2016)}. Inspired by the techniques \cite{Terhal(2000),Rungta(2003),Wang(2016)}, the $q$-concurrence $C_q(\rho_F)$ for these states will be derived by an extremization as follows.

\textbf{Lemma 2}. The $q$-concurrence for isotropic states $\rho_F$ on Hilbert space $\mathbb{C}^d\otimes \mathbb{C}^d$  $(d \geq2)$ is given by
\begin{eqnarray}
C_q(\rho_F)=co(\xi(F,q,d))
\end{eqnarray}
where $F\in(1/d, 1]$ and $co()$ denotes the largest convex function that is upper bounded by a given function $\xi(F,q,d)$ defined as
\begin{eqnarray}
\xi(F,q,d)=1-\gamma^{2q}-(d-1)\delta^{2q}
\end{eqnarray}
with $\gamma=\sqrt{F}/\sqrt{d}+\sqrt{(d-1)(1-F)}/\sqrt{d}$ and $\delta=\sqrt{F}/\sqrt{d}-\sqrt{1-F}/\sqrt{d(d-1)}$.

The evaluation is essentially algebraic and quite tedious, as shown in Appendix B. For convenience, take $q=2$ as an example. The $2$-concurrence of an two-qubit isotropic state $\rho_F$ is given by \cite{Rungta(2003)}:
\begin{eqnarray}
C_2(\rho_F)=
\left\{
\begin{aligned}
  &0,    && 0\leq F\leq\frac{1}{2}
  \\
  &\frac{(1-2F)^2}{2},    &&   \frac{1}{2}\leq F\leq1
\end{aligned}
\right.
\end{eqnarray}
Moreover, from Theorem 1 it implies $C_2(\rho_F)\geq(1-2F)^2/2=C_2(\rho_F)$. This implies that the upper bound gives exact value of the $2$-concurrence for this qubit systems. For arbitrary $d\geq3$, the $2$-concurrence $C_2(\rho_F)$ is given by
\begin{eqnarray}
C_2(\rho_F)=
\left\{
\begin{aligned}
  &0,    &&  F\leq\frac{1}{d}
  \\
  &\xi(F,2,d),      && \frac{1}{d}\leq F\leq\frac{4(d-1)}{d^2}
  \\
   &\frac{dF-d}{d-1}+\frac{d-1}{d},   && \frac{4(d-1)}{d^2}\leq F\leq1
\end{aligned}
\right.
\label{eqnq=2}
\end{eqnarray}
Moreover, it is shown that $\|\rho^{T_A}_F\|_1=\|{\cal{R}}(\rho_F)\|_1=dF$ for $F >1/d$ \cite{Vidal(2002),Rudolph (2005)}. From Theorem 1 we get that
\begin{eqnarray}
C_2(\rho_F)\geq \frac{(dF-1)^2}{d^2-d}
\label{eqnisotropi1}
\end{eqnarray}
For simplicity, $C_2(\rho_F)$ in Eq.(\ref{eqnq=2}) and its lower bounds in Eq.(\ref{eqnisotropi1}) are shown in Fig.1 with $3\leq d \leq 10$. Especially, it illustrates that the present lower bound is very close to the exact value of the $2$-concurrence, which shows of the tightness of our bounds.

\begin{figure}
\begin{center}
\resizebox{190pt}{145pt}{\includegraphics{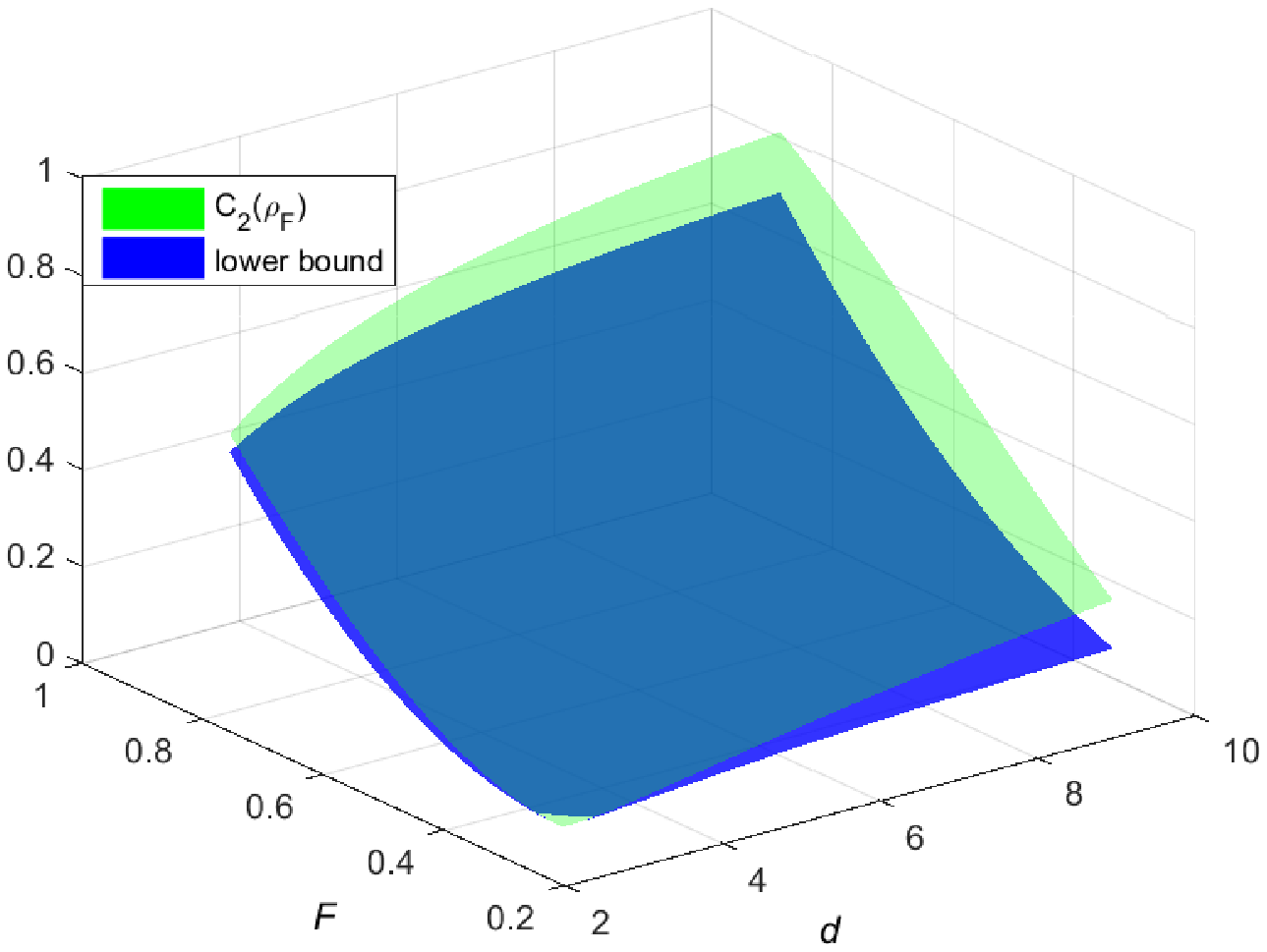}}

(a)

\resizebox{190pt}{145pt}{\includegraphics{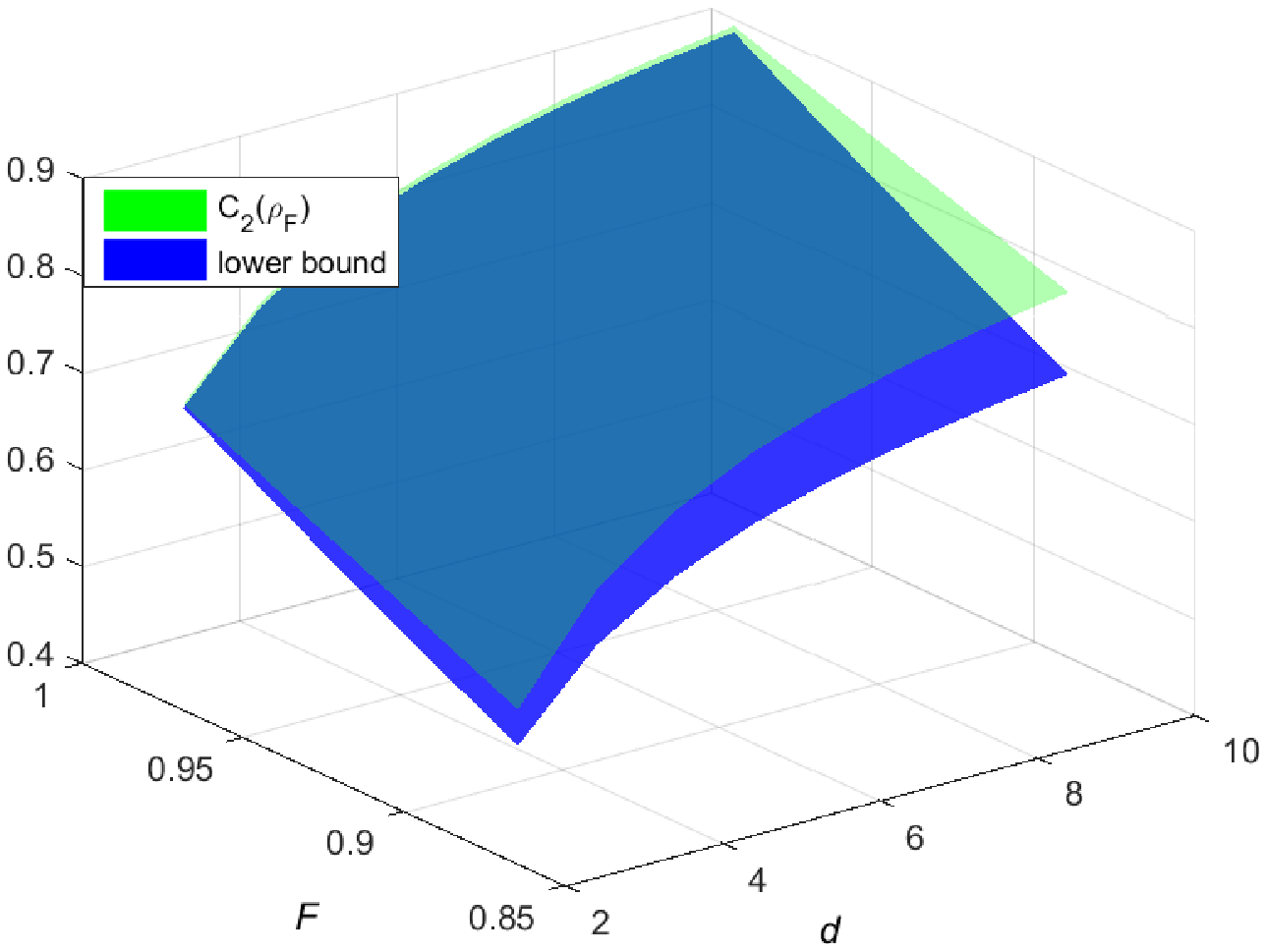}}

(b)
 \end{center}
\caption{\small (Color online) The $2$-concurrence (green) of isotropic states $\rho_F$ and lower bounds (blue) in Example 1.  (a) $3\leq d\leq 10$ and $1/d\leq F\leq 4(d-1)/d^2$. (b) $3\leq d\leq 10$ and $4(d-1)/d^2\leq F\leq 1$. }
 \label{Fig1}
\end{figure}

The lower bound in Theorem 1 can be used to detect the $q$-concurrence for all the entangled states of two-qubit or qubit-qutrit system because the PPT criterion is necessary and sufficient for the separability in both cases \cite{Peres(1996),Horodecki(1996)}. Unfortunately, it cannot detect all the other entangled states due to the limitation of the PPT criterion \cite{Peres(1996),Horodecki(1996)} and the realignment criterion \cite{Horodecki(1999),Rudolph (2005)}. Thus, it is intriguing to explore new bounds for the $q$-concurrence of general mixed states.

\section{The $q$-concurrence of superposition states}

Assume that a state $|\Gamma\rangle$ is generated by superposing two pure states $|\Phi\rangle$ and $|\Psi\rangle$, i.e.,   $|\Gamma\rangle=\alpha|\Phi\rangle+\beta|\Psi\rangle$. Our goal in this section is to explore the $q$-concurrence for these superposition states.  We discuss how the entanglement of superpositions of some given pure states is related to the entanglement contained in input states. In detail, we consider four cases: two component states in the superposition are bi-orthogonal states, one-sided orthogonal states, orthogonal states or arbitrary states.

\subsection{ Bi-orthogonal states}

\textbf{Definition 2}. Two bipartite states $|\Phi\rangle_{AB}$ and $|\Psi\rangle_{AB}$ on Hilbert space ${\cal H}_A\otimes {\cal H}_{B}$ are bi-orthogonal if they satisfy
\begin{eqnarray}
& &{\rm{Tr}}_B[{\rm{Tr}}_A(|\Phi\rangle\langle\Phi|){\rm{Tr}}_A(|\Psi\rangle\langle\Psi|)]=0
\label{eqside1}
\\
& &{\rm{Tr}}_A[{\rm{Tr}}_B(|\Phi\rangle\langle\Phi|){\rm{Tr}}_B(|\Psi\rangle\langle\Psi|)]=0
\label{eqside2}
\end{eqnarray}

For two bi-orthogonal states $|\Phi\rangle$ and $|\Psi\rangle$ we get up to local unitary transformations \cite{Gour(2007)1} that
\begin{eqnarray}
& &|\Phi\rangle=\sum^{d_1}_{i=1}a_i|i\rangle_A|i\rangle_B
\nonumber\\
& &|\Psi\rangle=\sum^{d}_{i=1}b_i|i+d_1\rangle_A|i+d_1\rangle_B
\label{eqnbio}
\end{eqnarray}
where $a_i, b_i$ are positive constants. In this case, the $q$-concurrence of the superposition state $|\Gamma\rangle$ will be evaluated as follows.

\textbf{Theorem 2}. Given two bi-orthogonal states $|\Phi\rangle$ and $|\Psi\rangle$, then the $q$-concurrence of the superposition $|\Gamma\rangle=\alpha|\Phi\rangle+\beta|\Psi\rangle$ satisfies
 \begin{eqnarray}
C_q(|\Gamma\rangle)=|\alpha|^{2q} C_q(|\Phi\rangle)+|\beta|^{2q} C_q(|\Psi\rangle)+h_q(|\alpha|^2)
\label{eqn1}
\end{eqnarray}
where $h_q(t)=1-t^q-(1-t)^q$, and $\alpha, \beta\in \mathbb{C}$ with $|\alpha|^2+|\beta|^2=1$.

\textbf{Proof}. From Eq.(\ref{eqnbio}) the reduced states of the system $A$ for $|\Phi\rangle$ and $|\Psi\rangle$ are diagonal in the same basis. More specifically, from Eq.(\ref{eqn0}), we have $C_q(|\Gamma\rangle)=F_q(\rho_A)$ with $\rho_A={\rm Tr}_{B}(|\Gamma\rangle\langle\Gamma|)$. It follows that the first $d_1$ eigenvalues of $\rho_A$ are given by $\{|\alpha|^{2}{a^2_i}, i=1, 2,\cdots, d_1\}$, and all the rest eigenvalues are given by $\{|\beta|^{2}{b^2_j}, j=1, 2,\cdots, d\}$. Thus, from Definition 1 we get
\begin{eqnarray}
C_q(|\Gamma\rangle)
&=&1-\sum_i(|\alpha|^{2}{a^2_i})^q-\sum_j(|\beta|^{2}{b^2_j})^q
\nonumber \\
&=&|\alpha|^{2q}(1-\sum_ia^{2q}_i)+|\beta|^{2q}(1-\sum_jb^{2q}_i)
\nonumber \\
& &+1-|\alpha|^{2q}-|\beta|^{2q}
\nonumber \\
&=&|\alpha|^{2q} C_q(|\Phi\rangle)+|\beta|^{2q} C_q(|\Psi\rangle)+h_q(|\alpha|^2)
\nonumber \\
\end{eqnarray}

From Lemma 1 for any density matrices $\rho$ and $\sigma$ we get the following inequalities:
\begin{eqnarray}
F_q(|\alpha|^2\rho+|\beta|^2\sigma)
   &\geq& |\alpha|^{2}F_q(\rho)+|\beta|^{2}F_q(\sigma)
\label{eqnT2}
  \\
F_q(|\alpha|^2\rho+|\beta|^2\sigma)
 &\leq& |\alpha|^{2q}F_q(\rho)+|\beta|^{2q}F_q(\sigma)
  \nonumber\\
& &+h_q(|\alpha|^2)
\label{eqnT1}
\end{eqnarray}
Moreover, from Lemma 1 the Eq.(\ref{eqnT1}) holds iff $\rho$ and $\sigma$ are orthogonal. Since $|\Phi\rangle$ and $|\Psi\rangle$ are bi-orthogonal, their reduced density matrices $\rho_A$ and $\sigma_A$ are orthogonal. Thus, from Eq.(\ref{eqnT1}) we get Eq.(\ref{eqn1}). $\square$

Note that the entanglement of the superposition is related to the average of the entanglement of two states being superposed. For convenience, define the increase of $q$-concurrence entanglement for the superposition state $|\Gamma\rangle=\alpha|\Phi\rangle+\beta|\Psi\rangle$ as follows:
\begin{eqnarray}
\Delta C_q(|\Gamma\rangle)=C_q(|\Gamma\rangle)-(|\alpha|^2C_q(|\Phi\rangle)
+|\beta|^2C_q(|\Psi\rangle))
\label{eqndeta1}
\end{eqnarray}
For the bi-orthogonal states $|\Phi\rangle$ and $|\Psi\rangle$, we obtain the following Corollary 1.

\textbf{Corollary 1}. Given two bi-orthogonal states $|\Phi\rangle$ and $|\Psi\rangle$,  the increase of the $q$-concurrence of the superposition state $|\Gamma\rangle$ is upper bounded by one ebit, i.e.,  $\Delta C_q(|\Gamma\rangle)\leq 1$.

\textbf{Proof}. From Theorem 2, we obtain
\begin{eqnarray}
C_q(|\Gamma\rangle)-[|\alpha|^{2q}C_q(|\Phi\rangle)+|\beta|^{2q}C_q(|\Psi\rangle)]
=h_q(|\alpha|^2)
\label{eqndeta2}
\end{eqnarray}
Moreover, since $|\alpha|, |\beta|\leq1$, it is obvious that
\begin{align}
\Delta C_q(|\Gamma\rangle)
=&C_q(|\Gamma\rangle)-(|\alpha|^{2}C_q(|\Phi\rangle)
  +|\beta|^{2}C_q(|\Psi\rangle))
\nonumber \\
 \leq &C_q(|\Gamma\rangle)-(|\alpha|^{2q}C_q(|\Phi\rangle)+|\beta|^{2q}C_q(|\Psi\rangle))
\label{eqndeta3}
\end{align}
Thereby, combing Eqs. (\ref{eqndeta2}) and (\ref{eqndeta3}) we get
\begin{eqnarray}
\Delta C_q(|\Gamma\rangle)\leq h_q(|\alpha|^2)\leq 1
\label{eqndeta4}
\end{eqnarray}
which implies that the increase of the $q$-concurrence for the superposition state $|\Gamma\rangle$  cannot be greater than one ebit. $\Box$

\textit{Example 2}. Consider the superposition state
\begin{eqnarray}
|\Gamma\rangle=\alpha|\Phi\rangle+\beta|\Psi\rangle
\label{eqnr11}
\end{eqnarray}
with $\alpha, \beta \in \mathbb{C}$  and $|\alpha|^2+|\beta|^2=1$, where $|\Phi\rangle=\cos\theta|00\rangle+\sin\theta|11\rangle$ and $|\Psi\rangle=\cos\phi|22\rangle+\sin\phi|33\rangle$ are generalized bipartite entangled states for $\theta,\phi\in (0, \pi/2)$. Note that $|\Phi\rangle$ and $|\Psi\rangle$ are bi-orthogonal states. From Eq.(\ref{eqn0}) it is easy to check that $C_q(|\Phi\rangle)=1-\cos^{2q}\theta-\sin^{2q}\theta$, $C_q(|\Psi\rangle)=1-\cos^{2q}\phi-\sin^{2q}\phi$ and
\begin{eqnarray}
 C_q(|\Gamma\rangle)
 &=&1-|\alpha|^{2q}(\cos^{2q}\theta+\sin^{2q}\theta)
 \nonumber\\
 & &-|\beta|^{2q}(\cos^{2q}\phi+\sin^{2q}\phi)
  \label{}
\end{eqnarray}
Moreover, we have $h_q(|\alpha|^2)=1-|\alpha|^{2q}-|\beta|^{2q}$. Thus, Eq.(\ref{eqn1}) holds for the superposition state $|\Gamma\rangle$ in Eq.(\ref{eqnr11}), which is consistent with Theorem 2.  From Eq.(\ref{eqndeta1}) we have
\begin{eqnarray}
\Delta C_q(|\Gamma\rangle)
&=&C_q(|\Gamma\rangle)-(|\alpha|^{2}C_q(|\Phi\rangle)
    +|\beta|^{2}C_q(|\Psi\rangle))
 \nonumber\\
&=&(|\alpha|^{2}-|\alpha|^{2q})(\cos^{2q}\theta+\sin^{2q}\theta)
 \nonumber\\
& &+(|\beta|^{2}-|\beta|^{2q})(\cos^{2q}\phi+\sin^{2q}\phi)
\label{eqndeta6}
\end{eqnarray}

To show the result in Corollary 1, we take the special case of $\alpha=\beta=1/\sqrt{2}$ and $q=4$. As shown in Fig.2, it is apparent that the function $\Delta C_q(|\Gamma\rangle)$ of $\theta$ and $\phi$ satisfies $\Delta C_q(|\Gamma\rangle)\leq1$.

\begin{figure}
\begin{center}
\resizebox{220pt}{180pt}{\includegraphics{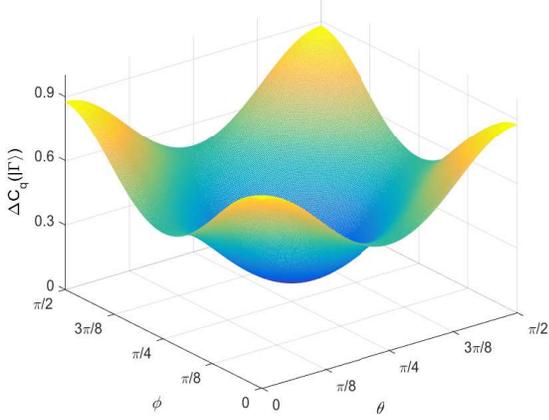}}
\end{center}
\caption{\small (Color online) The increase $\Delta C_q(|\Gamma\rangle)$  of the $q$-concurrence for the superposition state $|\Gamma\rangle$ in Example 2. Here, $\alpha=\beta=1/\sqrt{2}$ and $q=4$.}
\end{figure}

\subsection{One-side orthogonal states}

\textbf{Definition 3}. Two bipartite states $|\Phi\rangle_{AB}$ and $|\Psi\rangle_{AB}$ on Hilbert space ${\cal H}_A\otimes {\cal H}_{B}$ are one-side orthogonal if they satisfy only one of Eqs.(\ref{eqside1}) and (\ref{eqside2}).

Without loss of generality, we assume that two one-side orthogonal states satisfy Eq. (\ref{eqside1}). Up to local unitary transformations \cite{Gour(2007)1}, we have
\begin{eqnarray}
& &|\Phi\rangle=\sum^{d_1}_{i=1}a_i|i\rangle_A |i\rangle_B
 \nonumber \\
& & |\Psi\rangle=\sum^{d}_{i=1}b_i|i\rangle_A|i+d_1\rangle_B
\label{eqside3}
\end{eqnarray}
where $a_i$ and $b_i$ are positive constants.

Now, consider the case of $|\Phi\rangle$ and $|\Psi\rangle$ being one-sided orthogonal but not necessarily bi-orthogonal, i.e., they satisfy Eq. (\ref{eqside3}).

\textbf{Theorem 3}.  Given two one-side orthogonal states $|\Phi\rangle$ and  $|\Psi\rangle$, the $q$-concurrence of the superposition state $|\Gamma\rangle=\alpha|\Phi\rangle+\beta|\Psi\rangle$ is given by
\begin{eqnarray}
C_q(|\Gamma\rangle)
&=&|\alpha|^{2q}C_q(|\Phi\rangle)+|\beta|^{2q}C_q(|\Psi\rangle)
\nonumber\\
& &+h_q(|\alpha|^2)-|F_q(\rho_{A})-F_q(\rho_{B})|
\label{eqoneside1}
\end{eqnarray}
where $\rho_A$ and $\rho_B$ are reduced density matrices of $\rho_{AB}$, $\rho_{AB}:=|\alpha|^2|\Phi\rangle\langle\Phi|+|\beta|^2|\Psi\rangle\langle\Psi|$ with  $|\alpha|^2+|\beta|^2=1$, and $h_q$ is defined in Theorem 2.

\textbf{Proof}. Note that  $\||\Gamma\rangle\|=\sqrt{\langle\Gamma|\Gamma\rangle}=\sqrt{|\alpha|^2+|\beta|^2}=1$. It means that $|\Gamma\rangle$ is normalized. From Eq.(\ref{eqside3}) we know
\begin{eqnarray}
{\rm Tr}_{B}(|\Gamma\rangle\langle\Gamma|)
=|\alpha|^2{\rm Tr}_{B}(|\Phi\rangle\langle\Phi|)+|\beta|^{2}{\rm Tr}_{B}(|\Psi\rangle\langle\Psi|)
\label{eqnA}
\end{eqnarray}
which is also defined as the reduced density matrix $\rho_A$ by tracing $\rho_{AB}$ over the subsystem $B$. Hence, we get that ${\rm Tr}_{B}(|\Gamma\rangle\langle\Gamma|)=\rho_A$. From Eq.(\ref{eqn0}) it follows that
\begin{eqnarray}
C_q(|\Gamma\rangle)=F_q(\rho_A)
 \label{eqnA1}
\end{eqnarray}
Note that ${\rm Tr}_{A}(|\Gamma\rangle\langle\Gamma|)\neq\rho_B$ because $\rho_B$ is defined as the reduced density matrix by tracing $\rho_{AB}$ over the subsystem $A$, i.e.,
\begin{eqnarray}
\rho_B=|\alpha|^2{\rm Tr}_{A}(|\Phi\rangle\langle\Phi|)+|\beta|^{2}{\rm Tr}_{A}(|\Psi\rangle\langle\Psi|)
\label{eqnABB}
\end{eqnarray}
while
\begin{eqnarray}
{\rm Tr}_{A}(|\Gamma\rangle\langle\Gamma|)
 &=&|\alpha|^2{\rm Tr}_{A}(|\Phi\rangle\langle\Phi|)+\alpha\beta^\ast{\rm Tr}_{A}(|\Phi\rangle\langle\Psi|)
\nonumber\\
 & &+\alpha^\ast\beta{\rm Tr}_{A}(|\Psi\rangle\langle\Phi|)
\nonumber\\
&   &+|\beta|^{2}{\rm Tr}_{A}(|\Psi\rangle\langle\Psi|)
\end{eqnarray}
From Eqs.(\ref{eqside3}) and (\ref{eqnABB}) the first $d_1$ eigenvalues of $\rho_B$ are given by $\{|\alpha|^{2}{a^2_i}|i=1, 2,\cdots, d_1\}$, and all the rest are shown as $\{|\beta|^{2}{b^2_j}|j=1, 2,\cdots, d\}$. Thus, according to the definition of $F_q(\rho)$ in Eq.(\ref{eqnFq}) we get
\begin{eqnarray}
\!\!\!\!\!\!\!\!\!F_q(\rho_B)
&=&1-\sum_i(|\alpha|^{2}{a^2_i})^q-\sum_j(|\beta|^{2}{b^2_j})^q
\nonumber
\\
&=&|\alpha|^{2q}(1-\sum_ia^{2q}_i)+|\beta|^{2q}(1-\sum_jb^{2q}_i)
 \nonumber
\\
& &+1-|\alpha|^{2q}-|\beta|^{2q}
\nonumber\\
&=&|\alpha|^{2q} C_q(|\Phi\rangle)+|\beta|^{2q} C_q(|\Psi\rangle)
+h_q(|\alpha|^2)
 \label{eqnB1}
\end{eqnarray}

By utilizing Eqs.(\ref{eqnT1}) and (\ref{eqnA}) we have
 \begin{eqnarray}
 \!\!\!\!\!\!\!\!\!F_q(\rho_A)
 &\leq&|\alpha|^{2q}F_q({\rm Tr}_{B}(|\Phi\rangle\langle\Phi|))
 \nonumber \\
 & &+|\beta|^{2q}F_q({\rm Tr}_{B}(|\Psi\rangle\langle\Psi|))+h_q(|\alpha|^2)
 \nonumber\\
 &=&|\alpha|^{2q} C_q(|\Phi\rangle)+|\beta|^{2q} C_q(|\Psi\rangle)+h_q(|\alpha|^2)
 \label{eqnB2}
\end{eqnarray}
From Eqs.(\ref{eqnB1}) and (\ref{eqnB2}) $F_q(\rho_{B})\geq F_q(\rho_{A})$ for the one-side orthogonal states defined in Eq.(\ref{eqside1}). Due to Eqs.(\ref{eqnA1}) and (\ref{eqnB1}) we obtain
\begin{eqnarray}
 C_q(|\Gamma\rangle)
 &=&|\alpha|^{2q}C_q(|\Phi\rangle)+|\beta|^{2q}C_q(|\Psi\rangle)
\nonumber \\
 & &+h_q(|\alpha|^2)-(F_q(\rho_B)-F_q(\rho_A))
 \label{eqnAB1}
\end{eqnarray}

In a similar manner, for one-side orthogonal states in Eq.(\ref{eqside2}) we get that
\begin{eqnarray}
C_q(|\Gamma\rangle)
&=&|\alpha|^{2q}C_q(|\Phi\rangle)+|\beta|^{2q}C_q(|\Psi\rangle)
\nonumber \\
 & &+h_q(|\alpha|^2)-(F_q(\rho_{A})-F_q(\rho_{B}))
 \label{eqnAB2}
\end{eqnarray}
Combining Eqs.(\ref{eqnAB1}) and (\ref{eqnAB2}), we have completed the proof. $\square$

Note that $F_q(\rho_{A})=F_q(\rho_{B})$ holds for bi-orthogonal states $|\Phi\rangle$ and $|\Psi\rangle$. Thus we can obtain Eq.(\ref{eqn1}) for Theorem 2.  Moreover, since $|\Phi\rangle$ and $|\Psi\rangle$ are orthogonal pure states, we have $F_q(\rho_{AB})=h_q(|\alpha|^2)$. This implies $h_q(|\alpha|^2)\geq|F_q(\rho_{A})-F_q(\rho_{B})|$ from the triangle inequality of $F_q(\rho_{AB})$ in Lemma 1.

Similar to Corollary 1, we get the following result for one-side orthogonal states.

\textbf{Corollary 2}. Given two one-sided orthogonal states $|\Phi\rangle$ and $|\Psi\rangle$, the increase of the $q$-concurrence for the superposition state $|\Gamma\rangle$ is no more than one ebit, i.e.,  $\Delta C_q(|\Gamma\rangle)\leq1$.

\textbf{Proof}. From Theorem 3 we have
  \begin{eqnarray}
  \Delta C_q(|\Gamma\rangle)
&=&C_q(|\Gamma\rangle)-(|\alpha|^{2q}C_q(|\Phi\rangle)
  +|\beta|^{2q}C_q(|\Psi\rangle))
  \nonumber\\
&=&h_q(|\alpha|^2)-|F_q(\rho_{A})-F_q(\rho_{B})|
  \label{eqndeta04}
\end{eqnarray}
Note that
 \begin{eqnarray}
h_q(|\alpha|^2)-|F_q(\rho_{A})-F_q(\rho_{B}|\leq H_q(|\alpha|^2)
 \label{eqndeta4}
\end{eqnarray}
where the equality holds for the bi-orthogonal states $|\Phi\rangle$ and $|\Psi\rangle$.

Clearly, Eq.(\ref{eqndeta3}) holds for one-side orthogonal states $|\Phi\rangle$ and $|\Psi\rangle$. Thus, combining  Eqs.(\ref{eqndeta04}), (\ref{eqndeta4}) and  (\ref{eqndeta3}), we obtain $\Delta C_q(|\Gamma\rangle)\leq h_q(|\alpha|^2)\leq1$. This completes the proof. $\Box$

\emph{Example 3}. Consider the superposition state
\begin{eqnarray}
|\Gamma\rangle=\alpha|\Phi\rangle+\beta|\Psi\rangle
\label{eqnga1}
\end{eqnarray}
with $|\alpha|^2+|\beta|^2=1$, where $|\Phi\rangle=\cos\theta|00\rangle+\sin\theta|11\rangle$ and $|\Psi\rangle=\cos\phi|02\rangle+\sin\phi|13\rangle$ are one-side orthogonal entangled states for $\theta, \phi\in (0, \pi/2)$. From Eq.(\ref{eqn0}), it is easy to calculate that $C_q(|\Phi\rangle)=1-\cos^{2q}\theta-\sin^{2q}\theta$  $C_q(|\Psi\rangle)=1-\cos^{2q}\phi-\sin^{2q}\phi$, and
\begin{eqnarray}
C_q(|\Gamma\rangle)
 &=&1-(|\alpha|^2\cos^{2}\theta+|\beta|^2\cos^{2}\phi)^q
\nonumber\\
 & &-(|\alpha|^2\sin^{2}\theta+|\beta|^2\sin^{2}\phi)^q
\end{eqnarray}
According to Eq.(\ref{eqnFq}) one can check that
\begin{eqnarray}
F_q(\rho_{A})
&=&1-(|\alpha|^2\cos^{2}\theta+|\beta|^2\cos^{2}\phi)^q
\nonumber \\
 & &-(|\alpha|^2\sin^{2}\theta+|\beta|^2\sin^{2}\phi)^q
\end{eqnarray}
and
\begin{eqnarray}
F_q(\rho_{B})
&=&1-|\alpha|^{2q}\cos^{2q}\theta-|\alpha|^{2q}\sin^{2q}\theta
\nonumber\\
& &-|\beta|^{2q}\cos^{2q}\phi-|\beta|^{2q}\sin^{2q}\phi
\end{eqnarray}
Moreover, note that $h_q(|\alpha|^2)=1-|\alpha|^{2q}-|\beta|^{2q}$. It is easy to verify Eq.(\ref{eqoneside1}) for the superposition state $|\Gamma\rangle$ defined in Eq.(\ref{eqnga1}). This is consistent with Theorem 3.

Similar to Eq.(\ref{eqndeta6}), we can calculate $\Delta C_q(|\Gamma\rangle)$ of the superposition state $|\Gamma\rangle$ defined in Eq.(\ref{eqnga1}). For convenience, we take $\alpha=\beta=1/\sqrt{2}$ and $q=6$ for numerical evaluations. From Fig.3, it indicates that $\Delta C_q(|\Gamma\rangle)\leq 1$, which is consistent with Corollary 2.

\begin{figure}
\begin{center}
\resizebox{220pt}{170pt}{\includegraphics{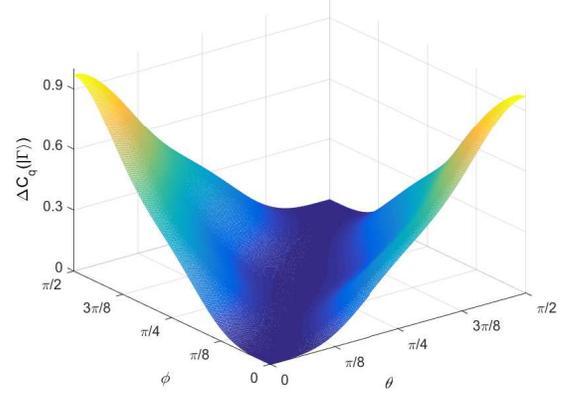}}
\end{center}
\caption{\small (Color online)The increase of the $q$-concurrence $\Delta C_q(|\Gamma\rangle)$ for the superposition state $|\Gamma\rangle$ in Example 3. Here, $\alpha=\beta=1/\sqrt{2}$ and $q=6$. }
\label{figure-3}
\end{figure}

\subsection{ Arbitrary states}

For general cases, two pure states that are superposed are not orthogonal. This means that the superposition state is not normalized. If we define $|\Gamma'\rangle=\frac{1}{c_+}|\Gamma\rangle$ with $|\Gamma\rangle=\alpha|\Phi\rangle+\beta|\Psi\rangle$ as the normalized state of $|\Gamma\rangle$, where $c_+$ is the normalization constant. By using Theorem 1, we prove the following inequality for its $q$-concurrence.

\textbf{Theorem 4}. Given any two different states $|\Phi\rangle$ and $|\Psi\rangle$, then the $q$-concurrence of the superposition $|\Gamma\rangle=\alpha|\Phi\rangle+\beta|\Psi\rangle$ satisfies
 \begin{eqnarray}
C_q(|\Gamma'\rangle)
&\leq& \frac{2}{c_+^2}(|\alpha|^{2q} C_q(|\Phi\rangle)+|\beta|^{2q} C_q(|\Psi\rangle)
\nonumber
\\
&&+h_q(|\alpha|^2)-|F_q(\rho_{A})-F_q(\rho_{B})|)
\label{eqnar}
\end{eqnarray}
or
\begin{eqnarray}
C_q(|\Gamma'_+\rangle)
&\leq &\frac{2}{c_+^2}(|\alpha|^{2q} C_q(|\Phi\rangle)+|\beta|^{2q}C_q(|\Psi\rangle)
\nonumber \\
&&+h_q(|\alpha|^2)-|F_q(\rho_{A})-F_q(\rho_{B})|)
\nonumber \\
&&
-\frac{c_-^2}{c_+^2}
\frac{(\|\sigma^{T_A}\|_1^{q-1}-1)^2}{m^{2q-2}-m^{q-1}}
\label{eqnar1}
\end{eqnarray}
where $|\alpha|^2+|\beta|^2=1$, and $\sigma$ is the density operator of $|\Gamma'_-\rangle$, i.e., $\sigma=|\Gamma'_-\rangle\langle\Gamma'_-|$. $\sigma^{T_A}$ stands for a partial transpose with respect to the subsystem $A$. $\|X\|_1$ denotes the trace norm.

\textbf{Proof}. By introducing an auxiliary state $|0\rangle_{a}$ and $|1\rangle_{a}$ on Hilbert space ${\cal H}_{a}$, consider the state
\begin{eqnarray}
|\Delta\rangle=\alpha|\Phi\rangle|0\rangle_{a}+\beta|\Psi\rangle|1\rangle_{a}
\label{}
\end{eqnarray}
The reduced state of $|\Delta\rangle$ is given by
\begin{eqnarray}
\rho_B=|\alpha|^2{\rm{Tr}}_A(|\Phi\rangle\langle\Phi|)
+|\beta|^2{\rm{Tr}}_A(|\Psi\rangle\langle\Psi|)
\label{eqnB11}
\end{eqnarray}
According to Eq.(\ref{eqnT1}), we get
\begin{eqnarray}
F_q(\rho_B)\leq|\alpha|^{2q} C_q(|\Phi\rangle)+|\beta|^{2q}  C_q(|\Psi\rangle)+h_q(|\alpha|^2)
\label{eqT2}
\end{eqnarray}

Similarly, for the reduced density matrix $\rho_A$ we have
\begin{eqnarray}
F_q(\rho_A)\leq|\alpha|^{2q} C_q(|\Phi\rangle)+|\beta|^{2q}  C_q(|\Psi\rangle)+h_q(|\alpha|^2)
\label{eqT20}
\end{eqnarray}
Thus, from Eqs.(\ref{eqT2}) and (\ref{eqT20}) we have
\begin{eqnarray}
\max\{F_q(\rho_A),F_q(\rho_B)\}
 &\leq &|\alpha|^{2q} C_q(|\Phi\rangle)+|\beta|^{2q}  C_q(|\Psi\rangle)
\nonumber\\
 & & +h_q(|\alpha|^2)
\label{eqT22}
\end{eqnarray}

Additionally, $\rho_B$ may be written into
\begin{eqnarray}
\!\!\!\!\!\!\rho_B&=&\frac{c^2_+}{2}{\rm{Tr}}_A[(\frac{\alpha|\Phi\rangle
+\beta|\Psi\rangle}{c_+})(\frac{\alpha^\ast\langle\Phi|+\beta^\ast\langle\Psi|}{c_+})]
\nonumber\\
&&
+\frac{c^2_-}{2}{\rm{Tr}}_A[(\frac{\alpha|\Phi\rangle-\beta|\Psi\rangle}{c_-})
(\frac{\alpha^\ast\langle\Phi|-\beta^\ast\langle\Psi|}{c_-})]
\label{eqnB22}
\end{eqnarray}
where $|\Gamma_\pm\rangle=\alpha|\Phi\rangle\pm\beta|\Psi\rangle$,  $|\Gamma'_\pm\rangle=\frac{1}{c_\pm}|\Gamma_\pm\rangle$ is the normalized state of $|\Gamma_\pm\rangle$, and $c_\pm$ are the normalization constants of $|\Gamma_\pm\rangle$.

Now, using Eqs.(\ref{eqnT2}) and (\ref{eqnB22}), we get the following inequalities:
\begin{eqnarray}
&&F_q(\rho_B)\geq\frac{c^2_+}{2}C_q(|\Gamma'_+\rangle)+\frac{c^2_-}{2}C_q(|\Gamma'_-\rangle)
\label{eqT1}
\\
&&F_q(\rho_A)\geq\frac{c^2_+}{2}C_q(|\Gamma'_+\rangle)
+\frac{c^2_-}{2}C_q(|\Gamma'_-\rangle)
\label{eqT10}
\end{eqnarray}
Similar to Eq.(\ref{eqT22}), Eqs.(\ref{eqT1}) and (\ref{eqT10})
can be written into:
\begin{align}
\!\!\!\!\!\!\min\{F_q(\rho_A),F_q(\rho_B)\}\geq\frac{c^2_+}{2}C_q(|\Gamma'_+\rangle)
+\frac{c^2_-}{2}C_q(|\Gamma'_-\rangle)
\label{eqT11}
\end{align}
If $\max\{F_q(\rho_A),F_q(\rho_B)\}=F_q(\rho_A)$, according to Eq.(\ref{eqT22}) we get
\begin{eqnarray}
F_q(\rho_B)+F_q(\rho_A)
 &\leq&|\alpha|^{2q} C_q(|\Phi\rangle)+|\beta|^{2q}  C_q(|\Psi\rangle)
  \nonumber\\
  &  &+h_q(|\alpha|^2)+F_q(\rho_B)
\label{eqT23}
\end{eqnarray}
It means that
\begin{eqnarray}
\min\{F_q(\rho_A),F_q(\rho_B)\}
&=& F_q(\rho_B)
\nonumber\\
&\leq &|\alpha|^{2q} C_q(|\Phi\rangle)+|\beta|^{2q}  C_q(|\Psi\rangle)
\nonumber\\
&   &+h_q(|\alpha|^2)-F_q(\rho_A)
\nonumber\\
&   &+F_q(\rho_B)
\label{eqT24}
\end{eqnarray}
If $\max\{F_q(\rho_A),F_q(\rho_B)\}=F_q(\rho_B)$, we can also get
\begin{eqnarray}
\min\{F_q(\rho_A),F_q(\rho_B)\}
&\leq &|\alpha|^{2q} C_q(|\Phi\rangle)+|\beta|^{2q}  C_q(|\Psi\rangle)
\nonumber \\
& &+h_q(|\alpha|^2)-F_q(\rho_B)
\nonumber\\
& &+F_q(\rho_A)
\label{eqT25}
\end{eqnarray}
Thus, combining Eqs.(\ref{eqT24}) and (\ref{eqT25}) we obtain
\begin{eqnarray}
\min\{F_q(\rho_A),F_q(\rho_B)\}
 &\leq &|\alpha|^{2q} C_q(|\Phi\rangle)+|\beta|^{2q}  C_q(|\Psi\rangle)
\nonumber\\
&&+h_q(|\alpha|^2)-|F_q(\rho_{A})
\nonumber\\
&   &-F_q(\rho_{B})|
\label{eqT3}
\end{eqnarray}
Moreover, from Eqs.(\ref{eqT11}) and (\ref{eqT3}) we have
\begin{align}
&\frac{c^2_+}{2}C_q(|\Gamma'_+\rangle)+\frac{c^2_-}{2}C_q(|\Gamma'_-\rangle)
\nonumber\\
&\leq|\alpha|^{2q} C_q(|\Phi\rangle)+|\beta|^{2q}C_q(|\Psi\rangle)+h_q(|\alpha|^2)
\nonumber \\
&\quad-|F_q(\rho_{A})-F_q(\rho_{B})|
\label{eqnT4}
\end{align}

If $C_q(|\Gamma'_-\rangle)=0$, i.e., the superposition state $|\Gamma'_-\rangle$ is separable, from Eq.(\ref{eqnT4}) it is obvious that
\begin{align}
C_q(|\Gamma'_+\rangle)
\leq &\frac{2}{c_+^2}(|\alpha|^{2q} C_q(|\Phi\rangle)+|\beta|^{2q}C_q(|\Psi\rangle)
\nonumber \\
& +h_q(|\alpha|^2)-|F_q(\rho_{A})-F_q(\rho_{B})|)
\label{}
\end{align}
If $C_q(|\Gamma'_-\rangle)>0$, i.e., a superposition state $|\Gamma'_-\rangle$ defined on $m\otimes n$($m\leq n$) systems is entangled, from Eq.(\ref{eqnTa}) we get
\begin{align}
C_q(|\Gamma'_+\rangle)
\leq &\frac{2}{c_+^2}(|\alpha|^{2q} C_q(|\Phi\rangle)+|\beta|^{2q}C_q(|\Psi\rangle)+h_q(|\alpha|^2)
\nonumber \\
&-|F_q(\rho_{A})-F_q(\rho_{B})|)
\nonumber \\
&
-\frac{c_-^2}{c_+^2}\frac{(\|\sigma^{T_A}\|_1^{q-1}-1)^2}{m^{2q-2}-m^{q-1}}
\label{}
\end{align}
where $\sigma$ is the density operator of $|\Gamma'_-\rangle$, i.e., $\sigma=|\Gamma'_-\rangle\langle\Gamma'_-|$. This completes the proof. $\square$

Corollaries 1 and 2 feature the maximal changes of the entanglement by the superposing two special states. For more general states, we get the following result.

\textbf{Corollary 3}. Given two arbitrary states $|\Phi\rangle$ and $|\Psi\rangle$, the increase of the $q$-concurrence for the superposition state $|\Gamma\rangle$ satisfies
\begin{eqnarray}
\nonumber&&C_q(|\Gamma'\rangle)-a (|\alpha|^{2}C_q(|\Phi\rangle)+|\beta|^{2} C_q(|\Psi\rangle))
\\&& \leq
a(h_q(|\alpha|^2)-|F_q(\rho_{A})-F_q(\rho_{B})|)
\label{eqnar00}
\end{eqnarray}
or
\begin{eqnarray}
&&C_q(|\Gamma'\rangle)-a(|\alpha|^{2} C_q(|\Phi\rangle)+|\beta|^{2} C_q(|\Psi\rangle))
\nonumber\\
& \leq& a(h_q(|\alpha|^2)-|F_q(\rho_{A})-F_q(\rho_{B})|)
\nonumber\\
&&-\frac{ac_-^2(\|\sigma^{T_A}\|_1^{q-1}-1)^2}{2m^{2q-2}-2m^{q-1}}
\label{eqnar11}
\end{eqnarray}
where $a=\frac{2}{c_+^2}$, and $\|\sigma^{T_A}\|_1$ is defined in Theorem 4.

\textbf{Proof}. Since $|\alpha|, |\beta|\leq1$, we get that
\begin{eqnarray}
|\alpha|^{2q}C_q(|\Phi\rangle)+|\beta|^{2q} C_q(|\Psi\rangle)&\leq& |\alpha|^{2} C_q(|\Phi\rangle)
\nonumber\\
&&+|\beta|^{2} C_q(|\Psi\rangle)
\end{eqnarray}
From Eqs.(\ref{eqnar}) and (\ref{eqnar1}), Corollary 3 is obtained by a straightforward evaluation. $\Box$

\emph{Example 4}. Consider the superposition state
\begin{eqnarray}
|\Gamma\rangle=\alpha|\Phi\rangle+\beta|\Psi\rangle
\label{eqnr02}
\end{eqnarray}
 with $\alpha, \beta \in \mathbb{C}$  and $|\alpha|^2+|\beta|^2=1$, where
\begin{eqnarray}
&&|\Phi\rangle=\cos\theta|00\rangle+\frac{1}{\sqrt{2}}\sin\theta|11\rangle
+\frac{1}{\sqrt{2}}\sin\theta|22\rangle
\nonumber\\
&&|\Psi\rangle=\cos\phi|03\rangle+\frac{1}{\sqrt{2}}\sin\phi|11\rangle
+\frac{1}{\sqrt{2}}\sin\phi|22\rangle
\end{eqnarray}
which are entangled for $\theta,\phi\in (0,\pi/2)$. From Eq.(\ref{eqn0}), we obtain
\begin{eqnarray}
&&C_q(|\Phi\rangle)=1-\cos^{2q}\theta-2^{1-q}\sin^{2q}\theta
\label{eqnfi0}
\\
&&C_q(|\Psi\rangle)=1-\cos^{2q}\phi-2^{1-q}\sin^{2q}\phi
 \label{eqnfi1}
 \end{eqnarray}
Here, we consider $\alpha=\beta=1/\sqrt{2}$ for the superposition state $|\Gamma\rangle$ in Eq.(\ref{eqnr02}). It should be clearly that
\begin{align}
&C_q(|\Gamma'_+\rangle)=1-\frac{2^q(\cos^{2}\theta+\cos^{2}\phi)^q
+2(\sin\theta+\sin\phi)^{2q}}{4^qc^{2q}_+}
\\
&C_q(|\Gamma'_-\rangle)=1-\frac{2^q(\cos^{2}\theta+\cos^{2}\phi)^q
+2(\sin\theta-\sin\phi)^{2q}}{4^qc^{2q}_-}
\label{eqnfi2}
\end{align}
where $|\Gamma'_\pm\rangle$ is the normalized state of $|\Gamma_\pm\rangle$ and $c_\pm=\sqrt{1\pm\sin\theta\sin\phi}$ is the normalization constant.  According to Eq.(\ref{eqnFq}), it is easy to check that
\begin{eqnarray}
F_q(\rho_A)=1-\frac{(\cos^{2}\theta+\cos^{2}\phi)^q}{2^q}
-\frac{(\sin^2\theta+\sin^2\phi)^q}{2^{2q-1}}
\label{eqnfi3}
\end{eqnarray}
and
\begin{eqnarray}
F_q(\rho_B)=1-\frac{\cos^{2q}\theta+\cos^{2q}\phi}{2^q}
-\frac{(\sin^2\theta+\sin^2\phi)^q}{2^{2q-1}}
\label{eqnfi4}
\end{eqnarray}
Moreover, $h_q(|\alpha|^2)=1-2^{1-q}$. Note that $C_q(|\Gamma'_-\rangle)=0$ iff $\theta=\phi$. We present the upper bound in Eq.(\ref{eqnar}) from Theorem 4 and the entanglement of superposition state $|\Gamma\rangle$ in Eq.(\ref{eqnr02}) in Fig.4. It indicates that the present bound is close to the exact value of the entanglement for superposition state $|\Gamma\rangle$. Howbeit, there may be some entanglement values of superposition states that can not be effectively evaluated, i.e., the present bound is larger than 1. Thus, the present bound in Eq.(\ref{eqnar}) may be further improved.
\begin{figure}
\begin{center}
\resizebox{220pt}{160pt}{\includegraphics{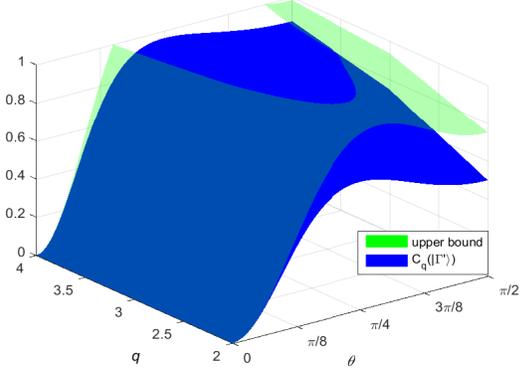}}
\end{center}
\caption{\small (Color online) The $q$-concurrence (blue) and upper bounds (green) of the superposition state $|\Gamma\rangle$ in Example 4. Here, $2\leq q\leq 4$, and $\theta\in(0, \pi/2)$. The upper bound of the $q$-concurrence is restricted to be no larger than 1.}
\end{figure}

When $C_q(|\Gamma'_-\rangle)>0$, i.e., $\theta\neq\phi$, for convenience, take $\theta=\pi/3$, $\phi=\pi/6$, and $q=2$ as an example. We get $\|\sigma^{T_A}\|_1=2.2571$.  Since the superposition state is a $3\otimes4$ system, which implies that $m=3$ in the right side of the inequality (\ref{eqnar1}). From  Eqs.(\ref{eqnfi0})-(\ref{eqnfi4}), a straightforward calculation shows the upper bound in the inequality (\ref{eqnar1}) being 0.8335, while $C_q(|\Gamma'_+\rangle)=0.6663$  according to Eq.(\ref{eqnar1}) in Theorem 4. This indicates that the upper bound in Eq.(\ref{eqnar1}) may be further improved. In a similar manner, it is easy to verify the validity of Corollary 3.

In Theorem 4, if two states $|\Phi\rangle$ and $|\Psi\rangle$ are orthogonal, i.e. $\langle\Phi|\Psi\rangle=0$, we have $c_\pm=1$. We can obtain the following Corollary from Theorem 4.

\textbf{Corollary 4}. Given two orthogonal states
$|\Phi\rangle$ and $|\Psi\rangle$ (not necessarily bi-orthogonal), the $q$-concurrence of the superposition state $|\Gamma\rangle=\alpha|\Phi\rangle+\beta|\Psi\rangle$ satisfies
\begin{eqnarray}
C_q(|\Gamma'\rangle)
\nonumber
&\leq&
 2(|\alpha|^{2q} C_q(|\Phi\rangle)+|\beta|^{2q} C_q(|\Psi\rangle)
\nonumber\\
&&+h_q(|\alpha|^2)-|F_q(\rho_{A})-F_q(\rho_{B})|)
\label{eqnnot}
\end{eqnarray}
or
\begin{align}
C_q(|\Gamma'_+\rangle)
\leq &2(|\alpha|^{2q} C_q(|\Phi\rangle)+|\beta|^{2q}C_q(|\Psi\rangle)+h_q(|\alpha|^2)
\nonumber \\
&-|F_q(\rho_{A})-F_q(\rho_{B})|)
\nonumber\\
&-\frac{(\|\sigma^{T_A}\|_1^{q-1}-1)^2}{m^{2q-2}-m^{q-1}}
\label{eqnr0}
\end{align}

\section{Conclusion}

Given an entanglement, how much is it entangled? The entanglement monotone has been introduced to solve this problem by quantifying the degree of entanglement. In this paper, inspired by general Tsallis entropy, we define a parameterized $q$-concurrence as a new entanglement monotone for any $q\geq 2$. We prove a lower bound of the $q$-concurrence for general states. The present bound is exact for two-qubit isotropic states. The new entanglement monotone is finally applied for characterizing the superposition state in terms of two states being superposed, especially for bi-orthogonal or one-sided orthogonal states. It shows that the increase of the $q$-concurrence for the superposition state is upper bounded by one ebit in both cases. These results are interesting in the entanglement theory, quantum information processing, quantum communication, and quantum many-body theory.

\section*{Acknowledgments}

This work was supported by the National Natural Science Foundation of China (No.61772437), Sichuan Youth Science and Technique Foundation (No.2017JQ0048), and Fundamental Research Funds for the Central Universities (No.2018GF07).


\appendix
\section{Proof of the Lemma 1}

\begin{itemize}
\item[(I)] Since the discrete spectra of $\rho_i$ of $\rho$ are in [0,1], we conclude the operator inequality $\rho^q\leq\rho$, where the equality holds iff $\rho$ is pure state. It follows that ${\rm{Tr}}\rho^q\leq 1$, where the equality holds iff $\rho$ is pure state. This implies $F_q(\rho)\geq0$, where the equality holds iff $\rho$ is pure state.
\item[(II)]  From the Schmidt decomposition in Eq.(\ref{eqnSchmidt1}), we know that the reduced density matrices $\rho_A$ and $\rho_B$ have the same spectra. From Eq.(\ref{eqnFq}), it is easy to show that $F_q(\rho_A)=F_q(\rho_B)$.
\item[(III)]  For $F_q(\rho_{AB})\leq F_q(\rho_A)+F_q(\rho_B)$, the key is an inequality (see Theorem 2 in \cite{Audenaert(2007)}) with the Schatten $q$-norm as
\begin{eqnarray}
1+\|\rho_{AB}\|^q_q\geq\|\rho_A\|^q_q+\|\rho_B\|^q_q
\label{}
\end{eqnarray}
This can be rewritten into
\begin{eqnarray}
{\rm{Tr}}\rho_A^q+{\rm{Tr}}\rho_B^q\leq1+{\rm{Tr}}\rho_{AB}^q
\label{}
\end{eqnarray}
which is equivalent to the inequality:
\begin{eqnarray}
1-{\rm{Tr}}\rho_{AB}^q\leq1-{\rm{Tr}}\rho_A^q+1-{\rm{Tr}}\rho_B^q
\label{}
\end{eqnarray}
It means that
\begin{eqnarray}
F_q(\rho_{AB})\leq F_q(\rho_A)+F_q(\rho_B)
\label{eqnAB01}
\end{eqnarray}
which completes the proof.

Similar to the von Neumann entropy, the subadditivity inequality leads to the triangle (or ``Araki-Lieb'') inequality \cite{AL}. For $|F_q(\rho_A)-F_q(\rho_B)|\leq F_q(\rho_{AB})$, the proof is inspired by its in ref.\cite{Rastegin(2011)}. Given a bipartite pure state $|\psi\rangle_{ABC}$, from the Schmidt decomposition of $|\psi\rangle_{ABC}$,  the density matrices $\rho_{AB}$ and $\rho_{C}$ have the same non-zero eigenvalues. Hence, $F_q(\rho_{AB})=F_q(\rho_{C})$. Similarly, we have $F_q(\rho_{A})=F_q(\rho_{BC})$. Combining these with the inequality (\ref{eqnAB01}), we get
\begin{eqnarray}
F_q(\rho_{A})-F_q(\rho_{B})\leq F_q(\rho_{AB})
\label{eqna-b1}
\end{eqnarray}
By symmetry, we also have
\begin{eqnarray}
F_q(\rho_{B})-F_q(\rho_{A})\leq F_q(\rho_{AB})
\label{eqna-b2}
\end{eqnarray}
Combining Eqs.(\ref{eqna-b1}) and (\ref{eqna-b2}), we have the claim that
\begin{eqnarray}
|F_q(\rho_A)-F_q(\rho_B)|\leq F_q(\rho_{AB})
\end{eqnarray}
\item[(IV)] For $\sum_i p_iF_q(\rho_i)\leq F_q(\sum_i p_i\rho_i)$, we firstly prove $\lambda F_q(\rho)+\mu F_q(\sigma)\leq F_q(\lambda\rho+\mu\sigma)$  with   $\lambda,\mu\geq0$ and $\lambda+\mu=1$. Here, from the Minkowski's inequality \cite{Lancaster(1985)} with positive semidefinite matrices $\rho$ and $\sigma$, we get
\begin{eqnarray}
({\rm{Tr}}(\rho+\sigma)^r)^{1/r}\leq({\rm{Tr}}\rho^r)^{1/r}+({\rm{Tr}}\sigma^r)^{1/r}
\label{eqnr0}
\end{eqnarray}
for $r\geq2$. From Eq.(\ref{eqnr0}), we have
\begin{eqnarray}
({\rm{Tr}}(\lambda\rho+\mu\sigma)^r)^{1/r}\leq\lambda({\rm{Tr}}\rho^r)^{1/r}
+\mu({\rm{Tr}}\sigma^r)^{1/r}
\label{eqnr1}
\end{eqnarray}
where $\lambda,\mu\geq0$ and $\lambda+\mu=1$. Due to $r\geq2$, from Eq.(\ref{eqnr1}),  we get
\begin{eqnarray}
{\rm{Tr}}(\lambda\rho+\mu\sigma)^r
&\leq&(\lambda({\rm{Tr}}\rho^r)^{1/r}+\mu({\rm{Tr}}\sigma^r)^{1/r})^r
\nonumber\\
&\leq&\lambda {\rm{Tr}}\rho^r+\mu {\rm{Tr}}\sigma^r
\label{eqnr2}
\end{eqnarray}
where the inequality (\ref{eqnr2}) is obtained from the convexity of the function $y=x^r$ for $r\geq2$. The inequality (\ref{eqnr2}) implies that
\begin{eqnarray}
\lambda(1-{\rm{Tr}}\rho^r)+\mu(1-{\rm{Tr}}\sigma^r)\leq1-{\rm{Tr}}(\lambda\rho+\mu\sigma)^r
\label{eqnr3}
\end{eqnarray}
By induction on $i$, we obtain the following inequality
\begin{eqnarray}
 \sum_i p_iF_q(\rho_i)\leq F_q(\sum_i p_i\rho_i)
 \label{eqnr4}
\end{eqnarray}
where $ \{p_i\}$ is probability distribution corresponding to density operators $\rho_{i}$ of $\rho$. The equality holds iff all the states $\rho_i$ are identical.

For $F_q(\sum_i p_i\rho_i)\leq\sum_i p^q_iF_q(\rho_i)+(1-\sum_i p^q_i)$, similar with Lemma 1 in \cite{Kim(2016)}. Suppose a joint state $\rho=\sum_i p_i\rho_{i}$, we get
\begin{eqnarray}
 F_q(\sum_i p_i\rho_i)&=&1-{\rm{Tr}}(\sum_i p_i\rho_i)^q
\nonumber\\
 &\leq& 1-\sum_i p^q_i{\rm{Tr}}(\rho^q_i)
\nonumber\\
&=&\sum_i p^q_i(1-{\rm{Tr}}(\rho^q_i))+1-\sum_i p^q_i
\nonumber\\
&=&\sum_i p^q_iF_q(\rho^q_i)+1-\sum_i p^q_i
\end{eqnarray}
The equality holds iff the states $\rho_i$ have support on orthogonal subspaces. The  proof is as follows: Let $\lambda_{ij}$ and $e_{ij}$  be the eigenvalues and corresponding eigenvectors of $\rho_i$. Note that $p_i\lambda_{ij}$ and $e_{ij}$ are the eigenvalues and eigenvectors of $\sum_i p_i\rho_i$. Thereby, we have
\begin{eqnarray}
F_q(\sum_i p_i\rho_i)&=&1-\sum_i (p_i\lambda_{ij})^q
\nonumber\\
&=&\sum_ip^q_i(1-\lambda^q_{ij})+1-\sum_i p^q_i
\nonumber\\
&=&\sum_ip^q_iF_q(\rho_i)+1-\sum_i p^q_i
\end{eqnarray}
which completes the proof.
\end{itemize}

\section{Proof of the Lemma 2}

The proof is inspired by recent techniques \cite{Vollbrecht(2001),Manne(2005), Wang(2016)} with local symmetry. The $q$-concurrence under the symmetry state $\rho_F$ is given by
\begin{eqnarray}
C_q(\rho_F)=co(\xi(F,q,d))
\end{eqnarray}
where the function $\xi(F,q,d)$ is defined as
 \begin{equation}
 \xi(F,q,d)=\inf\{C_q(|\psi\rangle)|f_{\Psi^{+}}(|\psi\rangle)=F,{\rm{rank}}(\rho_{\psi})\leq d\}
\end{equation}
where $\rm{rank}(\rho_{\psi})$ denotes the rank of the density operator $\rho_{\psi}=|\psi\rangle\langle\psi|$.

The $q$-concurrence of the pure state $|\psi\rangle=\sum^d_{i=1}\sqrt{\lambda_i}|a_ib_i\rangle$ is given in terms of the Schmidt coefficients by
\begin{eqnarray}
C_q(|\psi\rangle)=1-{\rm{Tr}}(\rho^{q}_A)=1-\sum^d_{i=1}\lambda^q_i
\label{eqnui}
\end{eqnarray}
In order to evaluate $f_{\Psi^{+}}(|\psi\rangle)$, we decompose $|\psi\rangle$ into its Schmidt decomposition as $|\psi\rangle=\sum^d_{i=1}\sqrt{\lambda_i}|a_ib_i\rangle=(U_A\otimes U_B)\sum^d_{i=1}\sqrt{\lambda_i}|ii\rangle$. From a straightforward calculation, we get $f_{\Psi^{+}}(|\psi\rangle)=\frac{1}{d}|\sum^d_{i=1}\sqrt{\lambda_i}v_{ii}|^2$ \cite{Terhal(2000)}, where $V=U^{\mathrm{T}}_AU_B$ and $v_{ij}=\langle i|V|j\rangle$.

Obviously, the value of $\xi(F,q,d)$ for $F\in(0,\frac{1}{d}]$ is easily obtained  by setting $\lambda_1=1, v_{11}=\sqrt{F}$, which yields $\xi(F,q,d)=0$. For  $F\in(\frac{1}{d},1]$, by using the Lagrange multipliers \cite{Rungta(2003)}, one can minimize Eq.(\ref{eqnui}) subject to the constraints
\begin{eqnarray}
&&\sum_{i}\lambda_i=1,
\\
&&\sum_{i}\sqrt{\lambda_i}=\sqrt{Fd}
\end{eqnarray}
with $Fd\geq1$. And then, the condition for an extremum is given by
\begin{eqnarray}
(\sqrt{\lambda_i})^{2q-1}+\mu_1\sqrt{\lambda_i}+\mu_2=0
\end{eqnarray}
where $\mu_1$ and $\mu_2$ denote the Lagrange multipliers. It is evident that $f(\sqrt{\lambda_i})=(\sqrt{\lambda_i})^{2q-1}$ is a convex function of $\sqrt{\lambda_i}$ for $q\geq2$.  Since a convex and a linear function cross each other in at most two points, this equation has maximally two possible nonzero solutions for $\sqrt{\lambda_i}$. Let $\gamma$ and $\delta$ denote these two positive solutions. The Schmidt vectors $\vec{\lambda}=\{\lambda_1,\lambda_2,\cdots,\lambda_d\}$ have coefficients
\begin{eqnarray}
\lambda_j=
\left\{
\begin{aligned}
  &\gamma^2, \,\,\,\,  j=1,\cdots,n &
  \\
  &\delta^2, \,\,\,\,  j=n+1,\cdots,n+m &
  \\
  &0,        \,\,\,\,  j=n+m+1,\cdots,d&
\end{aligned}
\right.
\end{eqnarray}
where $n+m\leq d$ and $n\geq1$. The minimization problem has been reduced into the following problem
\begin{eqnarray}
& &\mbox{ Given integers } n, m, n+m\leq d,
\nonumber\\
&\min & C_q(|\psi\rangle)
\label{eqminimum}
\\
&s.t.& n\gamma^2+m\delta^2=1,
\nonumber\\
& &n\gamma+m\delta=\sqrt{Fd}
\label{eqconstraint1}
\end{eqnarray}
where $C_q(|\psi\rangle)=1-n\gamma^{2q}-m\delta^{2q}$.

By solving Eq. (\ref{eqconstraint1}), we obtain
\begin{eqnarray}
\gamma^{\pm }_{nm}(F)=\frac{n\sqrt{Fd}\pm\sqrt{nm(n+m-Fd)}}{n(n+m)}
\label{eqconstraint02}
\end{eqnarray}
 and
\begin{eqnarray}
\delta^{\pm }_{nm}(F)&=&\frac{\sqrt{Fd}-n\gamma^{\pm }_{nm}}{m}
\nonumber\\
&=&\frac{m\sqrt{Fd}\mp\sqrt{nm(n+m-Fd)}}{m(n+m)}
\label{eqconstraint2}
\end{eqnarray}
Since $\gamma^{-}_{mn}=\delta^{+}_{nm}$, the function in Eq. (\ref{eqminimum}) has the same value for $\gamma^{+}_{nm}$ and $\gamma^{-}_{mn}$. Therefore, it only needs to  consider the solutions of $\gamma_{nm}:=\gamma^{+}_{nm}$. Since $\gamma_{nm}$ is a proper solution of Eq. (\ref{eqconstraint02}), the quantity inside the square root has to be nonnegative, which implies that $Fd\leq n+m$. On the other hand, $\delta_{nm}$ should be nonnegative in Eq. (\ref{eqconstraint2}), which implies that $Fd\geq n$. In this regime, one can verify that $\delta_{nm}(F)\leq\sqrt{Fd}/(n+m)\leq\gamma_{nm}(F)$. Note that $n=0$ is not defined. Hence, we have $n\geq1$.

To find the minimum of $C_q(|\psi\rangle)$ over all choices of $n$ and $m$, we can perform the minimization explicitly by regarding $n$ and $m$ as continuous variables.  It is completed by minimizing $C_q(|\psi\rangle)$ over the parallelogram defined by $1\leq n\leq Fd$ and $Fd\leq n+m\leq d$. Note that the parallelogram collapses to a line when $Fd=1$, i.e., the separability boundary. Within the parallelogram, we have $\gamma_{nm}\geq\delta_{nm}\geq0$.  $\gamma_{nm}=\delta_{nm}$ iff $n+m=Fd$ while $\delta_{nm}=0$ iff $n=Fd$. We first calculate the derivatives of $\gamma_{nm}$ and $\delta_{nm}$ with respect to $n$ and $m$ by differentiating the constraints (\ref{eqconstraint1}) as
\begin{eqnarray}
\nonumber&&\frac{\partial\gamma}{\partial n}=\frac{1}{2n}\frac{2\gamma\delta-\gamma^2}{\gamma-\delta}
\\
\nonumber &&\frac{\partial\delta}{\partial n}=-\frac{1}{2m}\frac{\gamma^2}{\gamma-\delta}
\\
\nonumber&&\frac{\partial\delta}{\partial m}=-\frac{1}{2m}\frac{2\gamma\delta-\gamma^2}{\gamma-\delta}
\\
&&\frac{\partial\gamma}{\partial m}=\frac{1}{2n}\frac{\delta^2}{\gamma-\delta}
\label{}
\end{eqnarray}
These can be used in Eq.(\ref{eqminimum}) to calculate the partial derivatives of $C_q(|\psi\rangle)$ with respect to $n$ and $m$ as
\begin{eqnarray}
\frac{\partial C_q}{\partial n}=(q-1)\gamma^{2q}-\frac{q\gamma^2\delta(\gamma^{2q-2}-\delta^{2q-2})}{\gamma-\delta}
\label{eqderivative1}
\end{eqnarray}
and
\begin{eqnarray}
\frac{\partial C_q}{\partial m}\nonumber&=&(q-1)\delta^{2q}-\frac{q\delta^2\gamma(\gamma^{2q-2}-\delta^{2q-2})}{\gamma-\delta}
\\&\leq&(q-1)\delta^{2q}-q\delta^2\gamma(\gamma+\delta)
\label{eqderivative21}
\\&\leq&(q-1)\delta^{2q}-2q\delta^{4}
\label{eqderivative22}
\\&\leq&(q-1-2q)\delta^{4}
\label{eqderivative23}
\\&\leq& 0
\label{eqderivative2}
\end{eqnarray}
where the inequality (\ref{eqderivative21}) is confirmed because $f(q)=(\gamma^{2q-2}-\delta^{2q-2})/(\gamma-\delta)$ is an increasing function of $q$, i.e.,
\begin{eqnarray}
 \frac{\partial f}{\partial q}=\frac{(2q-2)(\gamma^{2q-3}-\delta^{2q-3})}{\gamma-\delta}\geq0
 \label{eqnderivativef}
\end{eqnarray}
for $q\geq2$ and $\gamma\geq\delta$. The inequality (\ref{eqderivative22}) holds for $\gamma\geq\delta$. The inequality (\ref{eqderivative23}) is from that $\nu(\delta)=\delta^{2q}$ is a decreasing function of $q\geq2$. The inequality (\ref{eqderivative2}) is obtained for $2q\geq q-1$ with $q\geq2$.

Now we introduce two parameters $u=m-n$ and $v=m+n$, which correspond to motions parallel to and perpendicular to the $m+n=c$ ($c$ is a constant) boundaries of the parallelogram. The derivative of $C_q(|\psi\rangle)$ with respect to $u$ is given by
\begin{align}
\frac{\partial C_q}{\partial u}
\nonumber=&\frac{\partial C_q}{\partial n}\frac{\partial n}{\partial u}+\frac{\partial C_q}{\partial m}\frac{\partial m}{\partial u}
\\
\nonumber=&\frac{1}{2}(q-1)(\delta^{2q}-\gamma^{2q})
\\
\nonumber&-\frac{q(\gamma^{2q-2}-\delta^{2q-2})(\delta^2\gamma-\gamma^2\delta)}
{2(\gamma-\delta)}
\\ \leq &\frac{1}{2}(q-1)(\delta^{2q}-\gamma^{2q})-\frac{q}{2}(\gamma+\delta)\gamma\delta(\delta-\gamma)
\label{eqderivative30}
\\
\leq&\frac{1}{2}(q-1)(\delta^{4}-\gamma^{4})+\frac{q}{2}(\gamma^2-\delta^2)\gamma\delta
\label{eqderivative31}
\\
\leq&\frac{1}{2}[(\delta^{4}-\gamma^{4})+(\gamma^2-\delta^2)\gamma\delta]
\label{eqderivative32}
\\
\leq&-\frac{1}{2}[(\gamma^{2}-\delta^{2})(\gamma^2+\delta^2-2\gamma\delta)]
\label{eqderivative33}
\\
=&-\frac{1}{2}(\gamma+\delta)(\gamma-\delta)^3
\leq0
\label{eqderivative3}
\end{align}
where the inequality (\ref{eqderivative30}) holds for Eq.(\ref{eqnderivativef}). The inequality (\ref{eqderivative31}) is from that $g=(\delta^{2q}-\gamma^{2q})$ is a decreasing function of $q\geq2$, i.e.,
\begin{eqnarray}
\frac{\partial g}{\partial q}=2q(\delta^{2q-1}-\gamma^{2q-1})\leq0
\end{eqnarray}
Let  $h=\frac{1}{2}(q-1)(\delta^{4}-\gamma^{4}) +\frac{1}{2}q(\gamma^2-\delta^2)\gamma\delta$. We get
\begin{eqnarray}
\frac{\partial h}{\partial q}=\frac{-(\gamma^{2}-\delta^{2})(\gamma^2+\delta^2-\gamma\delta)}{2}\leq0
\end{eqnarray}
Thus $h$ is a decreasing function of $q\geq2$. The inequality (\ref{eqderivative32}) is achieved. The inequality (\ref{eqderivative33}) holds for  $\gamma\delta\leq 2\gamma\delta$.

From Eqs. (\ref{eqderivative2}) and  (\ref{eqderivative3}), it is obvious that $\frac{\partial C_q}{\partial m}\leq0$  within the parallelogram and $\frac{\partial C_q}{\partial u}\leq0$  except on the boundary $m+n=Fd$, where it is zero. These results imply that the minimum of  $C_q(|\psi\rangle)$ occurs at the vertex of $n=1$ and $m=d-1$. Thus, we get the minimum of $C_q(|\psi\rangle)$ as
\begin{eqnarray}
C_q(|\psi\rangle)=1-\gamma^{2q}_{1,d-1}-(d-1)\delta^{2q}_{1,d-1}
\end{eqnarray}
In this way, we derive an analytical expression of the function  $\xi(F,q,d)$  as
\begin{eqnarray}
\xi(F,q,d)=1-\gamma^{2q}-(d-1)\delta^{2q}
\label{eqnfqd}
\end{eqnarray}
where $\gamma$ and $\delta$ are defined as
\begin{eqnarray}
\nonumber&&\gamma=\frac{1}{\sqrt{d}}(\sqrt{F}+\sqrt{(d-1)(1-F)}) \\&&\delta=\frac{1}{\sqrt{d}}(\sqrt{F}-\frac{\sqrt{1-F}}{\sqrt{d-1}})
\label{}
\end{eqnarray}
Thus, the $q$-concurrence for isotropic states $C_q(\rho_F)=co(\xi(F,q,d))$, and $\xi(F,q,d)$ has the form in Eq.(\ref{eqnfqd}). This completes the proof of Lemma 2.

\end{document}